\documentclass[12pt,a4paper]{article}
\usepackage{cite}
\usepackage{epsfig}
\newcommand{\be}{\begin{equation}}
\newcommand{\ee}{\end{equation}}
\newcommand{\ba}{\begin{eqnarray}}
\newcommand{\ea}{\end{eqnarray}}

\newcommand{\btab}{\begin{tabular}}
\newcommand{\etab}{\end{tabular}}

\newcommand{\bit}{\begin{itemize}}
\newcommand{\eit}{\end{itemize}}

\def\spins{\hbox{\tiny spins}}
\def\real{\hbox{\tiny real}}

\def\PT{\hbox{\tiny PT}}
\def\NP{\hbox{\tiny NP}}

\def\SDG{\hbox{\tiny SDG}}
\def\NLL{\hbox{\tiny NLL}}
\def\bins{\hbox{\tiny bins}}
\def\DGE{\hbox{\tiny DGE}}

\catcode`\@=11 
\def \lsim {\mathrel{\mathpalette\@versim<}}
\def \gsim {\mathrel{\mathpalette\@versim>}}
\def\gappeq{\mathrel{\rlap {\raise.5ex\hbox{$>$}}{\lower.5ex\hbox{$\sim$}}}}
\def\lappeq{\mathrel{\rlap{\raise.5ex\hbox{$<$}}{\lower.5ex\hbox{$\sim$}}}}
\def\@versim#1#2{\vcenter{\offinterlineskip
\ialign{$\m@th#1\hfil##\hfil$\crcr#2\crcr\sim\crcr } }}
\catcode`\@=12 

\newcommand{\mycomm}[1]{\hfill\break $\phantom{a}$\kern-3.5em{\tt===$>$ \bf
#1}\hfill\break}
\newcommand{\mycommA}[1]{\hfill\break $\phantom{a}$\kern-3.5em{\tt   $>$ \bf
#1}\hfill\break}
\renewcommand{\thefootnote}{\fnsymbol{footnote}}
\newcommand{\ysl}{\mbox{$y$\hspace{-0.5em}\raisebox{0.1ex}{$/$}}}
\newcommand{\ksl}{\mbox{$k$\hspace{-0.5em}\raisebox{0.1ex}{$/$}}}
\newcommand{\psl}{\mbox{$p$\hspace{-0.4em}\raisebox{0.1ex}{$/$}}}
\newcommand{\nsl}{\mbox{$n$\hspace{-0.4em}\raisebox{0.1ex}{$/$}}}

\def\MSbar {\hbox{$\overline{\hbox{\tiny MS}}\,$}}

\def \as {\relax\ifmmode\alpha_s\else{$\alpha_s${ }}\fi}

\title{essai}
\author{}
\date{\today}

\def\beq{\begin{equation}}
\def\eeq{\end{equation}}

\begin{document}

\begin{titlepage}
\begin{flushright}
{\small CERN-TH/2002-371}\\
{\small UPRF-2002-19}\\
{\small December 2002}

\end{flushright}
\vspace{.3in}

\begin{center}
{\Large{\bf Heavy-Quark Fragmentation}}

\vspace{.8in}

{\large\bf Matteo Cacciari}$\;^{(1)}$\, and\, {\large\bf Einan Gardi}$\;^{(2,3)}$

\vspace{0.29in}

$^{(1)}$ Dipartimento di Fisica, Universit{\`a} di Parma, Italy,\\
and INFN, Sezione di Milano, Gruppo Collegato di Parma\\
\vspace*{10pt}
$^{(2)}$ TH Division, CERN, CH-1211 Geneva 23, Switzerland\\
\vspace*{10pt}
$^{(3)}$ Institut f{\"u}r Theoretische Physik, Universit{\"a}t Regensburg,\\
D-93040 Regensburg, Germany\\

\vspace{.4in}

\end{center}
\noindent  {\bf Abstract:}
We study perturbative and non-perturbative aspects of heavy-quark
fragmentation into hadrons, emphasizing the large-$x$ region, where~$x$
is the energy fraction of the detected hadron. We first prove that when
the moment index $N$ and the quark mass~$m$ get large simultaneously with the
ratio $N\Lambda/m$ fixed,   the fragmentation function depends on this
ratio alone. This opens up the way to formulate the non-perturbative
contribution to the fragmentation  function at large~$N$ as a shape
function of $m(1-x)$ which is convoluted with the Sudakov-resummed
perturbative result. We implement this resummation and the
parametrization of the corresponding shape function using Dressed Gluon
Exponentiation. The Sudakov exponent is calculated in a process
independent way from a generalized splitting function which describes
the emission probability of an off-shell gluon off a heavy quark.
Non-perturbative corrections are parametrized based on the renormalon
structure of the Sudakov exponent. They appear in moment space as an
exponential factor, with a leading contribution scaling as $N\Lambda/m$
and corrections of order $(N\Lambda/m)^3$ and higher. Finally, we
analyze in detail the case of $B$-meson production in $e^+e^-$
collisions, confronting the theoretical predictions with LEP
experimental data by fitting them in moment space.
\vspace{.25in}

\end{titlepage}

\def\thefootnote{\arabic{footnote}}
\setcounter{footnote} 0

\section{Introduction}

The fragmentation process, in which an energetic quark becomes a
hadron, is one of the most interesting processes in the physics of the
strong interaction, making an immediate link between the perturbative
regime and confinement. The case of heavy (bottom and, possibly, charm) 
quark fragmentation
is special in that the quark mass~$m$, being significantly larger than
the QCD scale~$\Lambda$, allows one  better theoretical control. In
particular, the mass provides a physical infrared cutoff for 
collinear radiation making the fragmentation process accessible to
perturbative methods.

The methodology to describe processes involving fragmentation is
based on factorization. In general ``factorization''  refers to
the separation of the physical process into subprocesses which are
mutually incoherent each depending on a separate scale. In
practice this term is mainly used in the context of separating the
perturbative and non-perturbative contributions~\cite{CS}. The
general procedure (see e.g.~\cite{Buras:qm}) is then to compute
the cross section perturbatively and complement it by a
convolution with a non-perturbative fragmentation
function\footnote{Note a possible source of confusion in the
terminology: the term ``fragmentation function'' is often used
both for the single-particle inclusive cross section in $e^+e^-$
annihilation and for the function describing the hadronization of
a parton into an observed hadron. In this paper we use the latter.
A fragmentation function is {\em not} an observable. When
evaluating the cross section for a given physical process, the
fragmentation function is convoluted with the proper coefficient
function.} describing the softening of the heavy-quark momentum as
it goes through the hadronization process.

Experimentally, observables involving bottom fragmentation are
particularly important. New experimental data have recently provoked
much debate~\cite{Cacciari:2002pa,Nason:2002ex,Cacciari:2002gn} about
the accuracy of predictions for $B$-meson hadroproduction  obtained by
convoluting the perturbative cross section for bottom quarks with
commonly used  phenomenological models~\cite{kart,peterson}.  There
exist many different viable implementations of a perturbative
calculation for heavy-quark production using massive or massless
quarks, different orders in~$\alpha_s$,  soft-gluon resummation,
Monte-Carlo simulations, etc. The phenomenology is particularly
intricate because the parameters of the non-perturbative fragmentation
function determined by fits to data are very sensitive to the details
of the perturbative description used.  Consequently the application of
these functions requires to use  the same kind of perturbative
description with which they were determined.  

On the theoretical side progress has been made at both the
perturbative~\cite{MN,Dokshitzer:fd,Dokshitzer:1995ev,Catani:2000ef,Keller:1998tf,CC} 
and the non-perturbative~\cite{JR,Webber_Nason} frontiers.
Nevertheless, the most pressing questions have not been answered, namely
\begin{itemize}
\item{} How should the non-perturbative fragmentation function be
parametrized, in particular in the region where the energy fraction
$x$ of the detected meson is large? This is
especially important since the distribution peaks at large~$x$.
\item{}
Is this function really universal? e.g. can the parameters be  fixed
from data on bottom fragmentation in $e^+e^-$ annihilation and  used in
various observables in hadron colliders?
\end{itemize}

Answering these questions requires deep insight into the
non-perturbative regime. However, a crucial step in addressing them is
making a clear and sensible separation between the perturbative and the
non-perturbative aspects and then dealing with both. Most previous work
on the subject is sharply divided between a purely perturbative
treatment, which concentrated on resumming the dominant perturbative
corrections~\cite{MN,Dokshitzer:1995ev,CC}, and a purely
non-perturbative approach, which disregarded the perturbative aspects.
The latter was either phenomenological in essence~\cite{kart,peterson},
or relied on general considerations such as the quark-mass
expansion~\cite{JR}, but avoided the large-$x$ region.
None of these approaches has led to a satisfactory description of the 
fragmentation function at large~$x$. This region is characterized by large
perturbative and non-perturbative effects, so both
aspects need to be addressed.

Owing to soft gluon radiation the perturbative result at any given
order diverges for $x \longrightarrow 1$. This makes Sudakov
resummation an essential ingredient in the description of heavy-quark
fragmentation~\cite{CC}. While important, Sudakov resummation (see e.g.
Fig.~\ref{fig:nlldgefull} below) does not lead, by itself, to a
phenomenologically acceptable description of the cross section: it is
much too hard and eventually becomes negative near $x=1$.
Non-perturbative corrections are required.

It is well known that due to renormalons~\cite{Beneke} the {\em
separation} between the perturbative and non-perturbative regimes
is arbitrary. This
means~\cite{Gardi:1999dq,Gardi:2001ny,Gardi:2002yg,Gardi:2002xm}
on the one hand that great care should be taken in controlling the
accuracy of the resummed perturbative expansion whenever
power-corrections are important but, on the other hand, that some
insight on non-perturbative physics can be gained based on
perturbation theory. Under the assumption that the renormalon
contribution dominates, this information can be used to
parametrize the non-perturbative corrections. Realizing this,
Nason and Webber~\cite{Webber_Nason} have used the structure of
infrared renormalons to determine the parametric form of the
leading power correction to the heavy-quark fragmentation function
in the large-$N$ limit, where $N$ is the moment (at large $N$, the
$N$-th moment is sensitive to $x\sim 1-1/N$). They found that it
should scale as~$N\Lambda /m$. This is an important result, which
we further establish and generalize here.

Based on the quark-mass expansion of~\cite{JR} and on a general
representation of the fragmentation function at large $N$ in terms of a
non-local lightcone matrix element at a large lightcone separation
(which follows directly from the definition~\cite{CS}),  we prove that
in the limit where $N$ and $m$ become large simultaneously, the
fragmentation function depends only on the ratio~$N\Lambda /m$ up to
corrections of order $1/N$.  This opens up the way to resum these
non-perturbative corrections into a  function of a single
variable~$N\Lambda /m$. The corresponding function  in momentum
fraction space depends on $m(1-x)$. This function  is convoluted with
the resummed perturbative distribution,  thereby introducing a shift
and a deformation of the shape by  non-perturbative effects. For this
reason it is called a ``shape function''. The scaling property of the
fragmentation function was known before (see~\cite{Buras:qm} are refs.
therein), but a formal proof based on the  properties of the hadronic
matrix element was never established.

Stepping forward from a leading, {\em additive} power correction to  a
shape function, which is a {\em multiplicative} correction in moment
space, is essential for describing differential cross sections near a
kinematic threshold~\cite{KS,Dokshitzer:1997ew,DIS,Gardi:2001ny}.
Shape-function based phenomenology has been developed particularly in
the context of event-shape distributions in $e^+e^-$
annihilation~\cite{KS,Dokshitzer:1997ew,Korchemsky:2000kp,Gardi:2001ny}
where it has proven very useful. The same methodology can be applied to
a large class of observables.

Our approach to the problem of heavy-quark fragmentation is based on
perturbation theory.  Instead of parametrizing the differential
distribution for the production of a heavy meson from a heavy quark, we
start off with a perturbative calculation of this distribution. The
result is then modified by power corrections through a convolution with
a shape function whose parametric form is deduced from the ambiguity in
the perturbative result.  A convolution of a perturbative (possibly
Sudakov-resummed) heavy-quark distribution with a phenomenological
non-perturbative function has been commonly used in describing
heavy-meson production. However,  there are deep conceptual differences
between our approach and the common practice: a) the resummation we
perform, which focuses on the effect of the running coupling in the
Sudakov exponent, is aimed at power accuracy, so that the separation
between the perturbative and non-perturbative components is power-like;
b) since the functional form of the non-perturbative component is
constructed according to the ambiguity of the perturbative one, it
should not be considered as a phenomenological model; c) since we only
rely on QCD, there is room for a systematic improvement of both the
perturbative and the non-perturbative ingredients.

We concentrate in this paper on the large-$x$ region, neglecting power
corrections which are suppressed by the energy of the quark,
${\cal O}(\alpha_s^2)$ perturbative corrections
 that are not logarithmically enhanced, as
well as power corrections in $\Lambda/m$ which are not enhanced by
the same power of $N$. Consequently, our description of the first
few moments is of limited accuracy.

Attempting to reach power accuracy at large $x$, our resummation
program is based on Dressed Gluon Exponentiation (DGE).
Calculating the Sudakov exponent as an integral over the running
coupling (a renormalon sum) we refrain from making a truncation at
some fixed logarithmic accuracy. The advantages of this
methodology have already been discussed in detail
in~\cite{Gardi:2001ny,Gardi:2002yg,Gardi:2002xm}. However, some
new features appear in the case of the fragmentation function
owing to the low scale $m/N$. As we will see, the perturbative
expansion of the Sudakov exponent reaches the minimal term already
at the next-to-leading-log (NLL) and a rather large difference
exists between truncation of the series at this order and a
principal-value (PV) regularization of the Borel integral. The
latter is advantageous as a basis for the parametrization of power
corrections, being it free of Landau type singularities and having a
smooth behavior at large~$N$.

Although data analysis is performed here only in one case, namely
the semi-inclusive cross section for $B$-meson production in
$e^+e^-$ annihilation at LEP, we are guided by the hypothesis that
the fragmentation function is universal, an idea that guided many
phenomenological applications in recent
years~\cite{Cacciari:1996wr,Cacciari:1997du,Nason:1999zj,Frixione:2002zv,Cacciari:2002pa}.
With universality in mind, we write our resummation formulae for
the cross section in a fully factorized form, where subprocesses
(such as the fragmentation function) depending on distinct
external scales appear as independent entities. This type of
factorization is not restricted to the resummed perturbation
theory, but instead it becomes a property of the full subprocess
including the non-perturbative contributions.

The paper is organized as follows. Section 2 recalls the standard
definition of the fragmentation function and then deals with the
heavy-quark fragmentation function in the large-$N$ limit on
general grounds. Section 3 summarizes a process-independent,
all-order calculation of the heavy-quark fragmentation function in
the large-$\beta_0$ limit. It is based on computing a generalized
splitting function describing an off-shell gluon emission off a
heavy quark. We then analyze the Borel singularity of the
fragmentation function and construct an ansatz for the shape
function. Section 4 specializes to the case of $e^+e^-$
annihilation. We begin by performing a renormalon calculation for
the differential cross section of the entire process and then
identifying the contributions which are enhanced near $x=1$.
Isolating the fragmentation subprocess we recover the result of
Sec.~3 and interpret the remaining subprocesses which form the
coefficient function.  In Sec.~5 we discuss the phenomenological
implications of our results and confront them with LEP data on
bottom production in $e^+e^-$ annihilation. Section 6 summarizes
our conclusions.

\section{Heavy quark fragmentation in the large-$x$ limit}

\begin{figure}[t]
\begin{center}
\epsfig{file=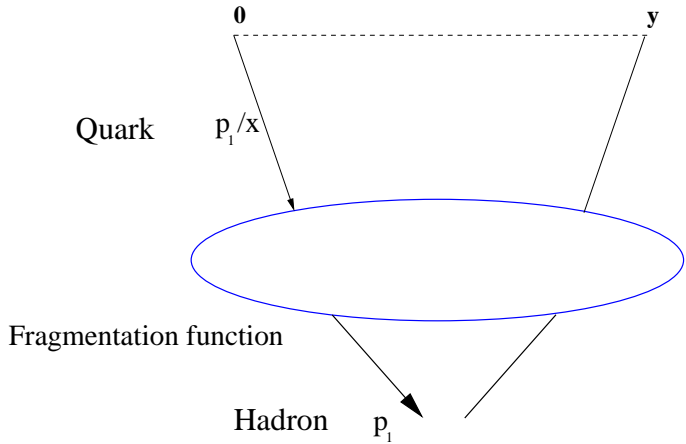,width=9cm}
\caption{\label{ffdef} The definition of the fragmentation function. 
The dashed line represents a path-ordered exponential.}
\end{center}
\end{figure}

The leading-twist quark fragmentation function, corresponding to
the probability of a quark to fragment into a hadron with a
longitudinal momentum fraction $0<x<1$ (see fig.~\ref{ffdef}), 
is defined\footnote{It
should be noted that, in contrast with $D(x;\mu^2)$, $F(p_1y;\mu^2)$ is 
Lorentz-frame dependent. Eqs.~(\ref{D_def}) and~(\ref{F_def})
refer to a Lorentz frame in which the decaying quark 
has a vanishing transverse momentum, in accordance with the 
Sudakov parametrization (\ref{Sudakov_par}) we will use below. 
It is for this reason that Eq.~(\ref{D_def}) differs by a factor $x^{d-2}$ 
(where $d$ is the space-time dimension) from the standard definition~\cite{CS}
corresponding to a frame where the hadron ($p_1$) has a vanishing
transverse momentum.} by~\cite{CS}, 
\be D(x;\mu^2)\equiv
\frac{1}{2\pi\,x}  \int_{-\infty}^{\infty} \frac{dy_{-}}{y_{-}}\,
\exp(i{p_1}y/x)\, F(p_1 y;\mu^2), \label{D_def} 
\ee where $F(p_1
y;\mu^2)$ stands for the following non-local matrix element \be
F(p_1 y;\mu^2) \equiv \frac{1}{4\, N_c}\,\sum_{X}\, {\rm
Tr}\left\{ \langle 0 \vert \ysl \Psi(y)\vert H(p_1)+X\rangle
\langle H(p_1)+X\vert \overline{\Psi}(0)\vert0\rangle_{\mu^2}
\right\}, \label{F_def} \ee renormalized at $\mu^2$, averaged over
spin and color of the initial quark and summed over all final
states containing a hadron $H$ with momentum $p_1$. We choose
light-cone coordinates such that $p_1$ has a large ``$+$''
component, while $y$ is a light-like vector in the ``$-$''
direction ($y_{+}, y_{\perp}\equiv~0$). Path-ordered exponential
factors are required to make~(\ref{F_def}) gauge invariant.
However, since we will eventually use the light-cone gauge $y\cdot
A=0$ where these factors become trivial, we will not write them
explicitly.

While this definition applies to fragmentation of both light and heavy
quarks, there is a fundamental difference in our ability to use the
theory in  the two cases. In the case of light quarks, only the $\mu^2$
evolution of the matrix element is calculable, whereas for heavy quarks
with mass $m\gg \Lambda$ the matrix element itself and thus
$D(x,m^2;\mu^2)$ can be calculated perturbatively, while
non-perturbative effects can be treated systematically as corrections.
The matrix element can be calculated since $m^2$ provides a natural
cutoff for collinear gluon radiation. It also provides a
natural scale for renormalization of the operator $\mu^2=m^2$, serving
as a factorization scale in a generic hard process. In the following we
will use simplified notation $D(x,m^2;\mu^2=m^2)=D(x,m^2)$ and
$F(p_1y,m^2;\mu^2=m^2)=F(p_1y,m^2)$.

By matching the general definition given above onto the heavy-quark effective
theory, Jaffe and Randall \cite{JR} have shown that in the large $m$ limit
the non-local matrix element in~(\ref{F_def}) can be written in terms of a
function of the product $(p_1 y)\, \bar{\Lambda}/m$ up to corrections
which are suppressed by $\bar{\Lambda}/m$:
\be
\frac{F(p_1 y, m^2)}{p_1 y}\,\longrightarrow \,\exp\left(-i{p_1}y\right)\,
\left[{\cal F}(p_1 y \,\bar{\Lambda}/m)+{\cal O}(\bar{\Lambda}/m)\right].
\label{JR_result}
\ee
Physically $\bar{\Lambda}$ appears as the energy of the light quark (including
the binding energy) in a meson, corresponding to $M-m$, where $M$ and $m$ are
the heavy meson and quark masses, respectively.
This particular dependence of $F(p_1 y,m^2)$, which follows directly from QCD,
has a phenomenological significance: it constrains possible models for the
parametrization of the heavy-quark fragmentation.
Due to the difficulty to distinguish between perturbative and non-perturbative
corrections at large~$x$, Jaffe and Randall exclusively concentrate in their
analysis on the first few Mellin moments of the $D(x,m^2)$,
\be
\tilde{D}(N,m^2) =\int_0^1 dx\; x^{N-1} D(x,m^2).
\label{D_mom}
\ee

As explained in the introduction, the behavior of the
fragmentation function in the large-$x$ limit is particularly
interesting both from a purely theoretical point of view and for
practical applications. As we show in the following sections, if
properly resummed, perturbation theory does provide a basis for
the analysis of heavy-quark fragmentation at large $x$,  so long
as $m (1-x) \gsim {\bar{\Lambda}}$. Of course, power corrections
can be ignored only if $m(1-x) \gg \bar{\Lambda}$, so in practice
the perturbative calculation must be supplemented by
non-perturbative corrections.

While phenomenological parameters are definitely needed for a precise
description of~(\ref{D_def}), it is important to constrain theoretically
its
functional form as much as possible. In the rest of this
section we show how the large-$x$ behaviour of $D(x,\mu^2)$ can be
constrained based on the Jaffe and Randall result (\ref{JR_result}).

First, let us rewrite the definition of the moments (\ref{D_mom})
for $N\gg 1$  as
\be
\tilde{D}(N,m^2) \simeq\int_{1-N_0/N}^1 dx \exp\left(-N(1-x)\right) D(x,m^2),
\label{D_N_mod}
\ee
where we used the fact that the integral is dominated by the
 $x \longrightarrow 1 $ region to expand $\ln x \simeq -(1-x)$ and to
 modify the lower integration bound to $1-N_0/N$ with $1 \ll N_0\ll N$.
Corrections of order $1/N$ are neglected here.
 A possible choice of $N_0$ is $\sqrt{N}$. This guarantees that the
 region $x\sim 1-1/N$ dominating the $N$-th moment (\ref{D_mom}) is
 fully included in~(\ref{D_N_mod}).

Substituting (\ref{D_def}) into (\ref{D_N_mod}) and changing the order of
integration we obtain,
\ba
\label{int_order}
\tilde{D}(N,m^2)&\simeq& \frac{1}{2\pi}\int_{-\infty}^{\infty} \,d(p_1 y)\,
\frac{F(p_1 y,m^2)}{p_1 y}\,\int_{1-N_0/N}^1\, \frac{dx}{x}  \,
\exp\left(-N(1-x)+i{p_1}y/x\right)\nonumber \\
&\simeq&\frac{1}{2\pi}\int_{-\infty}^{\infty} \,d(p_1 y)\,
\frac{F(p_1 y,m^2)}{p_1 y}\,\exp(ip_1y)\,\int_0^{N_0/N}\, d\xi  \,
\exp\left((-N+i{p_1}y)\,\xi\right),
\ea
where in the second line we used the large $x$ approximation and changed the
integration variable to $\xi\equiv 1-x$. Performing the $\xi$ integral we get
\[
\left[\exp\left(-N_0+ip_1y\,N_0/N\right)-1\right]\,\frac{\exp(ip_1y)}{ip_1y-N}.
\]
In the first factor we can safely take the limit $N\longrightarrow
\infty$ and then $N_0\longrightarrow \infty$ removing the $N_0$
dependence completely. The result is: \ba \label{int_f}
\tilde{D}(N,m^2)&\simeq& \frac{-1}{2\pi i}\int_{-\infty}^{\infty}
\,d(p_1 y)\, \frac{F(p_1 y,m^2)}{p_1 y}\,\frac{1}{p_1y+iN} \,
\exp\left(i{p_1}y\right), \ea

This last integral can be performed by closing a contour in the complex
$p_1 y$ plane, assuming that the matrix element ${F(p_1 y,m^2)}/{p_1 y}$
itself has no singularities. This integral should be done with
some care:  the phase of
${F(p_1 y,m^2)}/{p_1 y}$ dominates that of the exponential
factor, so the contour should be closed in the lower half plane.

To verify that the phase of the matrix element dominates, consider
the inverse Fourier transform to~(\ref{D_def}): \be \frac{F(p_1
y,m^2)}{p_1y}=  \int_{0}^{1} \frac{dx}{x}
\,{D(x,m^2)}\,\exp(-i{p_1}y/x)\, , \label{D_def_inverse} \ee where
we used the fact that $D(x,m^2)$ has support only between $0$ and
$1$. Now, since $D(x,m^2)$ is a positive definite function,
Eq.~(\ref{D_def_inverse}) represents a weighted average of the
phase factor $\exp(-i{p_1}y/x)$. This phase is opposite in sign
and larger than $i{p_1}y$ for any~\mbox{$0<x<1$}.

Having establish this, the integral (\ref{int_f}) can be evaluated
by closing the integration contour in the lower half plane (the
integral along the contour at infinity vanishes), picking up the
pole at $p_1y=-iN$. The result is the following remarkable
relation \ba \tilde{D}(N,m^2)&=& \left. \frac{F(p_1 y,m^2)}{p_1
y}\, \, \exp\left(i{p_1}y\right)\right\vert_{p_1y=-iN}+\,\,{\cal
O}(1/N). \label{large_N} \ea Eq. (\ref{large_N}) associates the
behavior of the fragmentation function at large $N$ and that of
the non-local matrix element~(\ref{F_def}), analytically continued
in the complex $p_1y$ plane and evaluated at asymptotically large
light-cone separation $p_1y=-iN$. We recall that similar
asymptotic relations exist in deep inelastic scattering between
light-cone matrix elements and moments of twist two~\cite{KM} and
twist four \cite{DIS} distributions.

Using now the result by Jaffe and Randall (\ref{JR_result}) we deduce that
\ba
\tilde{D}(N,m^2)&\simeq&
\left. \,{\cal F}(p_1 y \,\bar{\Lambda}/m) \,
\right\vert_{p_1y=-iN}+{\cal O}(1/N),
\label{large_N_JR}
\ea
namely, that in the limit where $m$ and $N$ get large
simultaneously~$\tilde{D}(N,m^2)$ becomes, asymptotically, a function of
a single parameter $N \bar{\Lambda}/m$, up to corrections suppressed by
$1/N$ or by $\bar{\Lambda}/m$. It should be noted that (\ref{JR_result})
holds at large $m$ for any $p_1y$ (not necessarily large), and likewise,
(\ref{large_N}) holds at large $N$ for general $m$. However, from the
two equations together it follows that there is one, specific way to
take the simultaneous limit getting a function of a single argument,
namely with the ratio $N/m$ fixed. 

We recall that Ref.~\cite{JR} used Eq.~(\ref{JR_result}) to derive a
scaling law for the fragmentation function in $x$ space. They also
performed moment space analysis but concluded that their result is only
applicable to the first few moments, due to the divergence of
the expansion in~$\bar{\Lambda}/m$. This problem is avoided, however, upon taking the limit where~$m$ and $N$ get large simultaneously.

The appearance of the scale $m/N$, or in $x$ space, $m(1-x)$ is not
surprising. As we discuss below,  this scale emerges naturally in
perturbation theory: $m(1-x)$ is the transverse momentum at which
soft-gluon radiation from the heavy quark peaks. It is directly related
to fact that the ``dead cone''~\cite{Dokshitzer:fd} angle is
proportional to the ratio between the quark mass and its energy (see
Sec.~3.2.1).

Let us also recall that the scaling of the fragmentation function at
large $x$ as $m(1-x)$ was discussed in the literature in the past. A
particularly intuitive argument in favour of this scaling law was
provided in~\cite{Buras:qm}. It was shown there that this behavior
follows from simple kinematic considerations if one assumes that the
differential cross section depends on the mass of the heavy quark and
on its energy only through the ratio, i.e. through its velocity.

The presence of a correction ${\cal O}(N\Lambda /m)$ was also predicted
by~\cite{Webber_Nason} based on a renormalon calculation in the
large-$N$ limit. In what follows we will adopt the assumption
of~\cite{Webber_Nason} concerning renormalon dominance at large~$N$. 
We will be using the renormalon structure of the Sudakov
exponent~\cite{Gardi:2001ny,DGE} to construct an ansatz for ${\cal
F}(p_1 y \,\bar{\Lambda}/m)$. As we explain in Sec.~3.2, by excluding
an additional logarithmic dependence on~$m$, Eq. (\ref{large_N_JR}) has
important consequences concerning the nature of the renormalon
singularities in the Sudakov exponent beyond the large-$\beta_0$ limit.

\section{Heavy-quark fragmentation function -- a process-independent
renormalon calculation}

\subsection{An off-shell splitting function}

Consider gluon emission off an outgoing heavy quark in a generic process.
Assume that the quark is on-shell after the emission
$p_1^2=m^2$, while the gluon has some time-like virtuality $k^2$.
\begin{figure}[htb]
\begin{center}
\centerline{{\epsfxsize5.0cm\epsfbox{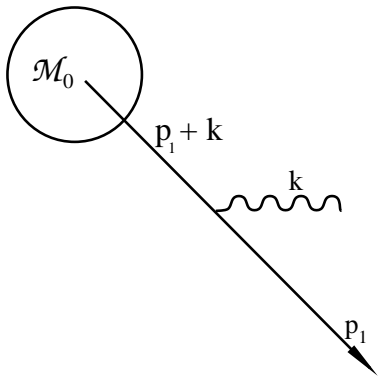}}}
\end{center}
\caption{Single gluon emission off a heavy quark involved in a generic hard
process~${\cal M}_0$.}
\label{SGE}
\vspace*{0pt}
\end{figure}

The amplitude for the emission of a single gluon off this quark
(Fig.~\ref{SGE}) is
\begin{equation}
\label{Amp}
{\cal M}=g_s\,t^a\,{\epsilon^{\lambda}_{\mu}}^* \,\frac{1}{(k+p_1)^2-m^2}\,
\bar{u}^{(s)}(p_1)
\gamma^\mu(\psl_1+\ksl+m){\cal M}_0,
\end{equation}
where $t^a$ is a colour matrix, ${\epsilon^{\lambda}_{\mu}}^* $ is
the gluon polarization vector and ${\cal M}_0$ represents the rest
of the process.

We wish to compute radiative corrections which are associated with
the singularity of the quark propagator
$1/[(p_1+k)^2-m^2]=1/(2p_1k+k^2)$.
These corrections become dominant for
a class of observables, including the fragmentation function.
Choosing the lightcone axial gauge
$n\cdot A=0$,
\be
d_{\mu\nu}\equiv \sum_{\lambda}{\epsilon_\nu^{\lambda}}^*
\epsilon_\mu^{\lambda}=-g_{\mu\nu}+\frac{k_{\mu}n_{\nu}+k_{\nu}n_{\mu}}{kn},
\label{d}
\ee
where $n$ is some\footnote{The vector $n$ is usually set parallel to
the direction of the other incoming or outgoing parton in ${\cal M}_0$,
however this is not essential.} lightlike vector ($n^2=0$), interference
terms are suppressed and the entire singularity is contained in the
amplitude~(\ref{Amp}) squared,
\begin{eqnarray}
\label{M2}
\hspace*{-14pt}
\sum_{\mbox {\rm {\spins}}} {\cal{M}}{\cal{M}}^\dagger =
\frac{C_F g_s^2}{(2p_1k+k^2)^2}\,d_{\mu\nu}
{\rm Tr}\left\{(\psl_1+\ksl+m)\gamma^\nu
(\psl_1+m) \gamma^\mu (\psl_1+\ksl+m){\cal M}_0{\bar{\cal M}_0}\right\}.
\end{eqnarray}
Here a sum over the quark and gluon polarizations was taken.

Proceeding with the evaluation of (\ref{M2}) we write
\begin{eqnarray}
\label{M22}
\sum_{\mbox {\rm {\spins}}} {\cal{M}}{\cal{M}}^\dagger =
\frac{C_F g_s^2}{(2p_1k+k^2)^2}\,\left[r_1+r_2\right],
\end{eqnarray}
where $r_{1,2}$ correspond to the two terms in the propagator (\ref{d}),
\begin{eqnarray}
r_1&=&-{\rm Tr}\left\{(\psl_1+\ksl+m)\gamma_\mu
(\psl_1+m) \gamma^\mu (\psl_1+\ksl+m){\cal M}_0{\bar{\cal M}_0}\right\}
\nonumber\\
r_2&=&\frac{1}{kn}\,{\rm Tr}\left\{(\psl_1+\ksl+m)\,\left[\nsl
(\psl_1+m) \ksl+\ksl
(\psl_1+m) \nsl\right] (\psl_1+\ksl+m){\cal M}_0{\bar{\cal M}_0}\right\}.
\label{r12}
\end{eqnarray}
After some algebra we get
\begin{eqnarray}
r_1&=&-4m\,[m^2+p_1k+k^2]\,{\rm Tr}({\cal M}_0{\bar{\cal M}_0})
-2\,[2m^2+k^2]\,{\rm Tr}(\psl_1{\cal M}_0{\bar{\cal M}_0}) \nonumber \\
&&+4\,[-m^2+p_1k]\,{\rm Tr}(\ksl{\cal M}_0{\bar{\cal M}_0}) \nonumber \\
r_2&=&\frac{2(2p_1k+k^2)}{kn}\,\left\{m\,[2p_1n+kn]
\,{\rm Tr}({\cal M}_0{\bar{\cal M}_0})\right.
+\,[2p_1n+kn]\,{\rm Tr}(\psl_1{\cal M}_0{\bar{\cal M}_0})\nonumber \\
&&\left.\hspace*{80pt}+\,p_1n \,{\rm Tr}(\ksl{\cal M}_0{\bar{\cal M}_0})
 -p_1k \,{\rm Tr}(\nsl{\cal M}_0{\bar{\cal M}_0}) \right\}.
\label{r12_}
\end{eqnarray}
Next we introduce the following Sudakov parametrization (adopting the
notation of \cite{CC}),
\begin{eqnarray}
\label{Sudakov_par}
p_1^{\nu}&=&zp^{\nu}-k_{\perp}^{\nu}+\frac{k_{\perp}^2+(1-z^2)m^2}{z\,(2pn)}
\,n^{\nu}\nonumber \\
k^{\nu}&=&(1-z)p^{\nu}+k_{\perp}^{\nu}+\frac{k^2+k_{\perp}^2-(1-z)^2m^2}{(1-z)
\,(2pn)}\,n^{\nu},
\end{eqnarray}
where $p^2=m^2$ and $k_{\perp}$ is orthogonal to both $n$ and $p$, with
${k_{\perp}}^{\nu}{k_{\perp}}_{\nu}=-k_{\perp}^2$. In these variables,
\begin{eqnarray}
&&\hspace*{-30pt}
r_1=-4m\,[m^2+p_1k+k^2]\,{\rm Tr}({\cal M}_0{\bar{\cal M}_0})
-2\,[2m^2+k^2-(1-z)(2p_1k+k^2)]
\,{\rm Tr}(\psl{\cal M}_0{\bar{\cal M}_0}) \nonumber \\
&&\hspace*{-30pt}
+2\,(2p_1k+k^2)\,{\rm Tr}(\ksl_{\perp}{\cal M}_0{\bar{\cal M}_0})
+\frac{(2p_1k+k^2)}{pn} \,[-2(2-z)m^2+z(2p_1k+k^2)]\,
{\rm Tr}(\nsl{\cal M}_0{\bar{\cal M}_0})\nonumber
\\
r_2&=&2(2p_1k+k^2)\,\left\{m\,\frac{1+z}{1-z}
\,{\rm Tr}({\cal M}_0{\bar{\cal M}_0})\right.
+\,\frac{2z}{1-z}\,{\rm Tr}(\psl{\cal M}_0{\bar{\cal M}_0}) \nonumber \\
&&\hspace*{120pt}\left.-\frac{1}{1-z} \,{\rm Tr}(\ksl{\cal M}_0{\bar{\cal M}_0})
 +\frac{m^2}{pn} \,{\rm Tr}(\nsl{\cal M}_0{\bar{\cal M}_0}) \right\}
\label{r12_Sud}
\end{eqnarray}
and finally,
\begin{eqnarray}
\label{M2_final}
&&\hspace*{-15pt}
\sum_{\mbox {\rm {\spins}}} {\cal{M}}{\cal{M}}^\dagger =
\left.\frac{C_F g_s^2}{(2p_1k+k^2)^2}\,\right\{
2m\left[\frac{2z}{1-z}(2p_1k+k^2)\,-\,(2m^2+k^2)\right]\,
{\rm Tr}({\cal M}_0{\bar{\cal M}_0})\nonumber \\
&&\hspace*{-15pt}
\,\,+\,\,
2\,\left[\frac{1+z^2}{1-z}(2p_1k+k^2)-(2m^2+k^2)\right]\,
{\rm Tr}(\psl{\cal M}_0{\bar{\cal M}_0})
-\frac{2z}{1-z}
\,(2p_1k+k^2)\,{\rm Tr}(\ksl_{\perp}{\cal M}_0{\bar{\cal M}_0})\nonumber \\
&&\hspace*{-15pt}
\,\,+\,\,\left.\frac{2p_1k+k^2}{pn} \,[z(2p_1k+k^2)-2(1-z)m^2]\,
{\rm Tr}(\nsl{\cal M}_0{\bar{\cal M}_0})
\right\}.
\end{eqnarray}

The quark propagator has the following singularity:
\be
\label{prop}
\frac{1}{(p_1+k)^2-m^2}=\frac{1}{2p_1k+k^2}=
\frac{z(1-z)}{zk^2+k_{\perp}^2+(1-z)^2m^2}.
\ee
Therefore, the relevant limit\footnote{This is a generalization of
the quasi-collinear limit \cite{Catani:2000ef} to the case of an off-shell
gluon.} is the one in which $m^2$, $k^2_{\perp}$,  $k^2$,
all become small simultaneously, whereas the ratios between them
(which depend on the quark longitudinal momentum fraction
$z=(p_1n)/(p_1n+kn)$) are fixed.

In this limit the dominant contribution to the squared matrix
element~(\ref{M2_final}) is the one proportional to
${\rm Tr}(\psl{\cal M}_0{\bar{\cal M}_0})$. All the other terms are
suppressed by one of the small parameters. We thus proceed with the
following approximation,
\begin{eqnarray}
\label{M2_dominant}
\sum_{\mbox {\rm {\spins}}} {\cal{M}}{\cal{M}}^\dagger
\simeq\frac{8\pi\, C_F\alpha_s }{(2p_1k+k^2)^2}\,
\left[\frac{1+z^2}{1-z}(2p_1k+k^2)-(2m^2+k^2)\right]\,
{\rm Tr}(\psl{\cal M}_0{\bar{\cal M}_0}),
\end{eqnarray}
which is factorized into a product of the squared matrix element of the non-radiative
process ${\rm Tr}(\psl{\cal M}_0{\bar{\cal M}_0})$ and a splitting
function describing the probability of a single gluon emission. This
splitting function corresponds to the emission of an {\em off-shell gluon}
off a {\em heavy quark}, and thus it generalizes both~\cite{CC} where the
gluon is on-shell and~\cite{DGE} where the quark is massless.

\subsection{The quark fragmentation function in the large-$\beta_0$ limit}

Based on the general definition (\ref{D_def}) and the off-shell
splitting function derived above, we can now calculate the quark
fragmentation function in a process independent way. The calculation is
performed with a single {\em dressed} gluon, so it is exact at order
$\alpha_s$ and contains the leading contribution in the large-$\beta_0$
limit at higher orders. Using DGE, this result will be used to compute
the Sudakov exponent.

\subsubsection{NLO calculation}

Let us begin by extracting the Born-level (${\cal O}(\alpha_s^0)$) result.
Saturating the state $\vert H(p_1) +X\rangle$ by the outgoing on-shell quark
with momentum $p_1$ alone, and using the definition~(\ref{F_def}) one finds
that \be
F(p_1y)=(p_1y)\,\exp(-i p_1y).
\label{F_bl}
\ee
The $y_{-}$ integration in (\ref{D_def}) can be readily performed,
\[
\frac{1}{x}\,\frac{1}{2\pi}\,\int_{-\infty}^{\infty}\,d(p_1y)
{\rm e}^{i p_1y\left(\frac1x-1\right)}=\delta (1-x),
\]
identifying the longitudinal momentum fraction $x$ as $1$.

The NLO~(${\cal O}(\alpha_s)$) calculation proceeds in a similar way. The
state~$\vert H(p_1) +X\rangle$ is saturated by the outgoing on-shell quark
$p_1$ and off-shell gluon~$k$. Calculating the squared matrix
element~(\ref{F_def}) we get~(\ref{M2_dominant}) with
${\cal M}_0{\bar{\cal M}_0}=N_c\, \ysl\,\exp(-i p_1y/z)$, so
\be
F(p_1y)=\int d\phi\, \frac{p_1y}{z}\,{\rm e}^{{-ip_1y}/{z}}
\,\frac{8\pi\, C_F\alpha_s }{(2p_1k+k^2)^2}\,
\left[\frac{1+z^2}{1-z}(2p_1k+k^2)-(2m^2+k^2)\right]
\label{F_bl2}
\ee
where $d\phi$ is the gluon phase space with the longitudinal momentum of the ``detected'' quark $p_1n$ fixed.
We see that the state $\vert H(p_1) +X\rangle$  is effectively replaced by a
single outgoing on-shell quark with momentum~$p_1/z$, times some factors
which are incorporated into the splitting function.

Let us perform this integration with a fixed gluon virtuality.
The phase-space integral is
\be
\int d\phi=\int\frac{d^4k}{(2\pi)^3}\,\delta(2k_{+}k_{-}-k_{\perp}^2-k^2)=
\frac{1}{16\pi^2}\,\int\frac{dz}{z(1-z)}\,\int dk_{\perp}^2.
\ee
The contribution of a single real gluon emission to the quark fragmentation
function is therefore
\begin{eqnarray}
\label{non-integrated_form}
\hspace*{-80pt}&&D^{\real}(x,k^2/m^2)=
\frac{ C_F\alpha_s}{2\pi}\,\frac1{x(1-x)}\,
\int dk_{\perp}^2\,
\left[\frac{1+x^2}{1-x}
\frac{1}{2p_1k+k^2}-\frac{(2m^2+k^2)}{(2p_1k+k^2)^2}\right] =\\ \nonumber
&& \hspace*{-10pt}
\frac{ C_F\alpha_s}{2\pi}\,
\int dk_{\perp}^2\,
\left[\frac{1+x^2}{1-x}\,
\frac{1}{{xk^2+k_{\perp}^2+(1-x)^2m^2}} \,-\,\,x(1-x)\,
\frac{(2m^2+k^2)}{({xk^2+k_{\perp}^2+(1-x)^2m^2})^2}\right],
\end{eqnarray}
where we substituted the expression for the propagator~(\ref{prop}).

Next, performing the integration over~$k_{\perp}^2$ from $k_{\perp}^2=0$
(ignoring terms from the upper integration limit where the propagator is
non-singular), we get
\begin{eqnarray}
D^{\real}(x,k^2/m^2)
&=& -\,\frac{ C_F\alpha_s}{2\pi}\,
\left[\frac{1+x^2}{1-x}\,
\ln{\left(x\,(k^2/m^2)+(1-x)^2\right)}\right. \nonumber \\
 &&\left.\hspace*{80pt} \,+\,\,x(1-x)\,
 \frac{2+(k^2/m^2)}{{x\,(k^2/m^2)+(1-x)^2}}\right].
\end{eqnarray}
Setting $k^2\to 0$ in this expression we recover Eq.~(58)
of~\cite{CC}, as expected.

Let us return now to the non-integrated form of the real-emission
matrix element~(\ref{non-integrated_form}) and examine the
distribution of the radiation in some Lorentz frame where the
longitudinal momentum of the quark $pn=p_{+}\equiv \sqrt{q^2}/\sqrt{2}$ is
large: $q^2\gg k^2, k_{\perp}^2,m^2$ ($\sqrt{q^2}$~can be, for example, the centre-of-mass energy in $e^+e^-$ annihilation). The energies of the outgoing
quark and gluon are \ba \label{Ep1}
E_{p_1}&=&\frac{q^2x^2+k^2+m^2}{2x\sqrt{q^2}}\,\simeq\, \frac{\sqrt{q^2}}2\, x
+{\cal O}\left(1/\sqrt{q^2}\right)
\nonumber \\
E_{k}&=&\frac{q^2(1-x)^2+k^2+k_{\perp}^2}{2(1-x)\sqrt{q^2}} \,\simeq\,
\frac{\sqrt{q^2}}2\, (1-x) +{\cal O}\left(1/\sqrt{q^2}\right), 
\ea so in this frame the
longitudinal momentum fraction coincides with the energy fraction
and the emission angle is \be
\sin^2\theta=\frac{4k_{\perp}^2}{q^2x^2(1-x)^2}\,+\,{\cal
O}(1/q^4). \ee The leading (log-enhanced) contribution at ${\cal
O}(\alpha_s)$ is obtained by taking $k^2=0$ and $1-x\ll 1$, \be
\frac{dD^{\real}}{dx \,d\sin^2\theta}=C_F\,\frac{
\alpha_s}{\pi}\,\frac{\sin^2\theta}{(1-x)(\sin^2\theta+4m^2/q^2)^2}.
\ee This result implies that the radiation peaks at angles
$\theta\simeq 2m/\sqrt{q^2}$ and that it is depleted in the forward
direction ($\theta=0$). The authors of Ref.~\cite{Dokshitzer:fd}
called this depletion the ``dead cone'', contrasting it with the
enhanced radiation for $\theta\longrightarrow 0$ in the case of a
massless quark. In a boost-invariant formulation the radiation
peak appears at $|k_{\perp}| \simeq m(1-x)$. We have already seen
in Sec.~2 that $m(1-x)\sim m/N$ is the relevant scale at large
$x$. It will emerge again as the typical gluon virtuality ($k^2$)
and eventually, as the natural scale for the coupling in the
framework of the renormalon calculation that follows.

\subsubsection{All-order calculation in the large-$\beta_0$ limit}

\begin{figure}[t]
\begin{center}
\epsfig{file=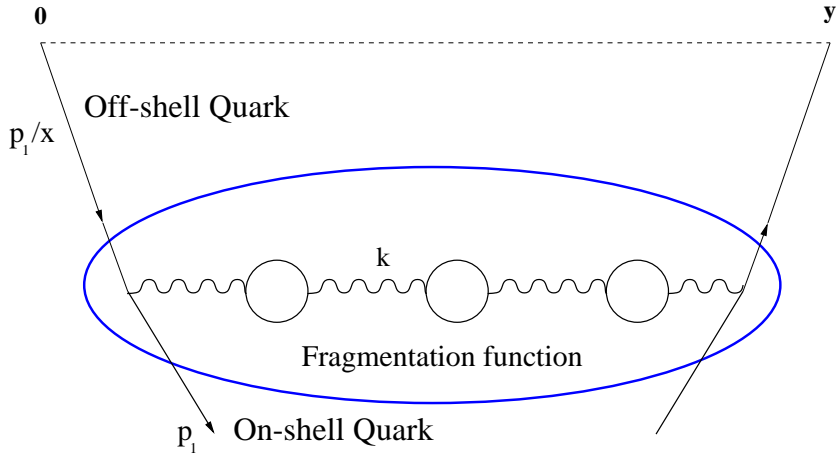,width=11cm}
\caption{\label{fig:renchain} The diagram contributing to
$\tilde{D}(N,m^2)$ in the light-cone axial gauge $A\cdot y=0$ and in the
large $N_f$ limit.}
\end{center}
\end{figure}

The ${\cal O}(\alpha_s)$ result for an off-shell gluon emission
derived in the previous section can be promoted to a single
dressed gluon (SDG) renormalon sum by integrating over the gluon
virtuality~$k^2$, with the coupling being renormalized at~$k^2$ (see
fig.~\ref{fig:renchain}).
It is convenient to define a factorization scheme and scale
invariant quantity, by taking the logarithmic derivative with
respect to~$m^2$. Contrary to the fragmentation function itself,
its logarithmic derivative is free of ultraviolet divergence. The
result, written in moment space~(\ref{D_mom}) as a scheme
invariant Borel transform~\cite{Grunberg}, is
\begin{eqnarray}
\label{ren_sum}
&&\hspace*{-30pt}
\frac{d \tilde{D}_{\SDG}(N,m^2)}{d\ln m^2}
= \frac{ C_F}{\beta_0}\,\int_0^{\infty}du\, \,T(u)\,
{\left(m^2/\Lambda^2\right)}^{-u}\,
\int_{0}^1\,dx\,\left(x^{N-1}-1\right)\,B_D(u,x),
\end{eqnarray}
where $x^{N-1}$ has been replaced by $x^{N-1}-1$ to account for virtual
corrections, and
\begin{eqnarray}
\label{ren_sum_B}
B_D(u,x)=\frac{-u\sin \pi u}{2\pi}\,{\rm e}^{cu}\,m^{2u}\,&& \\ \nonumber
&&\hspace*{-150pt}
\int_0^{\infty}\frac{dk^2}{k^{2(1+u)}}\,\left[\frac{1+x^2}{1-x}\,
\ln{\left(x\,(k^2/m^2)+(1-x)^2\right)} \,+\,x(1-x)\,
\frac{2+(k^2/m^2)}{{x\,(k^2/m^2)+(1-x)^2}}\right].
\end{eqnarray}
Note that in (\ref{ren_sum_B}) $m^2$ can be scaled out, so
$B_D(u,x)$ is independent of any scale and the dependence of
${d\ln \tilde{D}_{\SDG}(N,m^2)}/{d\ln m^2}$ on $m^2$ appears only
through ${\left(m^2/\Lambda^2\right)}^{-u}$. Here the factor
$(-u)$ is associated with the logarithmic derivative, the sine
appears upon taking the time-like discontinuity of the dressed
gluon propagator, and $T(u)$ is defined as the Laplace transform
of the coupling \be \label{borelcoupling}
A(k^2)=\beta_0{\alpha}_s(k^2)/\pi=\int_0^{\infty}\,du\,T(u)
\left(k^2/\Lambda^2\right)^{-u}. \ee $T(u)$ depends only on the
coefficients of the renormalization group equation of
$\alpha_s/\pi$. For a one-loop running coupling $T(u)=1$ and in
the two-loop case \be \label{A_B}
T(u)=(u\delta)^{u\delta}\exp(-u\delta)/\Gamma (1+u\delta), \ee
where $\delta\equiv \beta_1/\beta_0^2$, with $\beta_0 = \pi b_0$
and $\beta_1 = \pi^2 b_1$ as defined in Eq.~(28) of~\cite{CC}.
Defining $\Lambda$ in the ${\overline{\rm MS}}$ scheme, \be
\label{c}
c=5/3+{\cal O}(1/\beta_0).
\ee

Performing the integration over the gluon virtuality in~(\ref{ren_sum_B})
we get
\begin{eqnarray}
\label{x_space_result}
B_D(u,x)=-\,{\rm e}^{cu}\,\left(\frac{x}{(1-x)^2}\right)^u\,
\left[\frac{x}{1-x}\,(1-u)+\frac12(1-x)\,(1+u)\right].
\end{eqnarray}
Here we used integrals of the form
\[
\int_0^{\infty}\frac{dy}{y^{1+u}}\,\frac{1}{y+(1-x)^2}=
-\frac{\pi}{\sin \pi u}\,(1-x)^{-2(1+u)}
\]
with $y=x k^2/m^2$. Note that this $\sin \pi u$ factor in the
denominator cancels against the one in the numerator of
(\ref{ren_sum_B}), so there are no renormalon singularities
in~(\ref{x_space_result}). There is however, a convergence
constraint for the Borel integral~(\ref{ren_sum}) at
$u\longrightarrow \infty$. In the case of one-loop running
coupling ($T(u)=1$) the constraint is: \be
m^2(1-x)^2/x>\Lambda^2\exp(c). \label{1loop_converg_constraint}
\ee As we will see below, this singular behavior for $x
\longrightarrow 1$ translates into renormalons upon performing the
$x$ integration.

Note also the appearance of the factor $(1-u)$ in front of the
$x\longrightarrow 1$ singular term in~(\ref{x_space_result}). This
reflects a relation between the two terms in the squared matrix
element~(\ref{M2_dominant}). Let us examine the two corresponding terms
in the square brackets in~(\ref{ren_sum_B}): the first is the source of
the $1$ and the second, which can be represented as a logarithmic
derivative of the first with respect to $k^2$, yields the $-u$. As we
will see, in moment space this structure leads to the absence of a
renormalon singularity at  $u=1$, while renormalons do appear at all
other integer and half integer values.

Let us perform now the integration over $x$ in
Eq.~(\ref{ren_sum}). We obtain
\begin{eqnarray}
\label{N_ren_sum}
&&\hspace*{-30pt}
\frac{d\ln \tilde{D}_{\SDG}(N,m^2)}{d\ln m^2}
= \frac{ C_F}{\beta_0}\,\int_0^{\infty}du\, \,T(u)\,{\left(m^2/\Lambda^2\right)}
^{-u}\,B_{\tilde{D}}(u,N),
\end{eqnarray}
with
\begin{eqnarray}
\label{B_D}
B_{\tilde{D}}(u,N)&=&\,
\int_{0}^1\,dx\,\left(x^{N-1}-1\right)\,B_D(u,x)=\nonumber \\
&&\hspace*{-40pt} -{\rm e}^{cu} \,\left[ (1-u)\Gamma(-2u)
\left(\frac{\Gamma(N+1+u)}{\Gamma(N+1-u)}-\frac{\Gamma(2+u)}{\Gamma(2-u)}
\right)\right.\nonumber \\
&& \hspace*{50pt}\left.+
\frac12(1+u)\Gamma(2-2u)\left(\frac{\Gamma(N+u)}{\Gamma(N+2-u)}-
\frac{\Gamma(1+u)}{\Gamma(3-u)}\right)\right].
\end{eqnarray}
As anticipated, $B_{\tilde{D}}(u,N)$ has infrared renormalon singularities
(simple poles) at all positive integer and half integer values of $u$,
with the single exception of $u=1$.

\subsubsection{Ultraviolet subtraction for the fragmentation function}

In order to obtain the fragmentation function itself
Eq.~(\ref{N_ren_sum}) needs to be integrated. The integration
involves a subtraction of the ultraviolet divergent contribution,
as standardly done in collinear factorization. The result takes
the form:
\begin{eqnarray}
\label{N_space_result}
&& \hspace*{-20pt}
 \tilde{D}_{\SDG}(N,m^2;\mu^2)\,=\, 1\,+\nonumber \\
&& \,\frac{ C_F}{\beta_0}\,\int_0^{\infty}\,du \,T(u)\,
\left(m^2/{\Lambda}^{2}\right)^{-u}\,\left[ -\frac{B_{\tilde{D}}(u,N)}{u}\,+\,
\left(\mu^2/m^2\right)^{-u}\,\frac{B_{\tilde{E}}(u,N)}{u}\right],
\end{eqnarray}
where~$\mu$ is a factorization scale and~$B_{\tilde{E}}(u,N)$ is the
Altarelli-Parisi evolution kernel (the subscript $E$ stands for
{\em Evolution}),
\be
\label{B_E}
B_{\tilde{E}}(u,N)\,=\,\sum_{n=0}^{\infty}\frac{\gamma_n(N)\,u^{n}}{n!}.
\ee
This subtraction guarantees the existence of the integral near $u=0$
by canceling the $1/u$ singularity.

We stress that this subtraction is required only because we calculate
the fragmentation process alone. In any infrared and collinear safe
observable the coefficient function will compensate the $1/u$
singularity. This is demonstrated in Sec.~4 in the case of the
differential cross section in $e^+e^-$ annihilation. Through the
evolution term the fragmentation function becomes factorization-scheme
and scale dependent although the full result for the cross section is
not.

The leading order coefficient in~(\ref{B_E}) is renormalization-scheme
invariant, and it equals to the~$u=0$ limit of~(\ref{B_D}), ensuring the
cancellation of the $1/u$ singularity in~(\ref{N_space_result}),
\[
\gamma_0(N)=S_1(N)-\frac{3}{4}+\frac12\left(\frac{1}{N+1}-\frac{1}{N}\right),
\]
where $S_k(N)\equiv \sum_{j=1}^N {1}/{j^k}$, so that
$S_1(N)=\Psi(N+1)+\gamma_E$.
In the $\overline{\rm MS}$ factorization scheme~$\gamma_n(N)$ are known to
two-loop order~($n=1$) in full~\cite{FRS} and to all orders in the
large-$\beta_0$ limit~\cite{Gracey}.
The latter is given by\footnote{Note
that using the scheme invariant Borel transform beyond the large-$\beta_0$
limit, the relation between the coefficients of $B_{\tilde{E}}(u,N)$ and
those of $\tilde{E}(N,A)$ becomes non trivial (see e.g.~\cite{Gardi:2002xm}).}
\begin{eqnarray}
\label{E}
\tilde{E}(N,A) &=&\sum_{n=0}^{\infty}
 \gamma_n(N) A^{n+1}={\cal A}\,\bigg[ \Psi(N+A)-\Psi(1+A)  \\ \nonumber
&+&\frac{N-1}{2}
\left(\frac{A^2+2A-1}{1+A}\,\frac{1}{N+A}-\frac{(1+A)^2}{2+A}\frac{1}{N+1+A}
\right) \bigg]\,+\,{\cal O}(1/\beta_0)
\end{eqnarray}
with
\be
{\cal A} = \frac{\sin \pi A}{\pi}\,\,\frac{\Gamma(4+2A)}{6\Gamma(2+A)^2}\,+
\,{\cal O}(1/\beta_0).
\label{G_large_Nf}
\ee
${\cal A}$ is the all-order large-$\beta_0$ expression in~$\overline{\rm MS}$
for the gluon bremsstrahlung effective charge~\cite{CMW} or the cusp anomalous
dimension~\cite{KR,KM,Beneke:1995pq}, and
$A=\beta_0\alpha_s/\pi$ is the large-$\beta_0$ coupling in $\overline{\rm MS}$.

Using eqs.~(\ref{B_D}) through~(\ref{E}) it is straightforward to extract the
large-$\beta_0$ coefficients of the heavy-quark fragmentation
function~$\tilde{D}(N,m^2)$ in the $\overline{\rm MS}$ factorization scheme
to all orders,~for example
\begin{eqnarray}
\label{large_beta0_exp}
&& \tilde{D}_{\SDG}(N,m^2)\,\,=\,1\,+\frac{C_F}{\beta_0}
\left\{\bigg[-S_1^2(N)+\right.
\nonumber \\
&&
\left(\frac{1}{N}-\frac{1}{N+1}+1\right) S_1(N)+1 -S_2(N)
+\frac{1}{2(N+1)}-\frac{1}{2N}-\frac{1}{(N+1)^2}\bigg] A(m^2)\,+
\nonumber \\
&&\hspace*{5pt}
\bigg[-\frac{2}{3} S_1^3(N)+\left(\frac1N-\frac1{N+1}-\frac23\right) S_1^2(N)
 \nonumber\\
&&+\left(\frac19-\frac{\pi^2}{3}
+\frac{8}{3}\frac{1}{N}-\frac{8}{3}\frac{1}{N+1}-\frac{1}{(N+1)^2}-
\frac{1}{N^2}\right) S_1(N)\nonumber \\
&&\hspace*{5pt}- \frac{5}{6} S_2(N)+\frac{1}{6} S_3(N)+\left(\frac14-\frac16
\frac{1}{N+1}+\frac{1}{6}\frac{1}{N}\right) \pi^2+\frac{173}{96}+\frac{19}{36}
\frac{1}{N}-\frac{19}{36} \frac{1}{N+1}\nonumber \\
&&\hspace*{5pt}\left.
-\frac{5}{12} \frac{1}{N^2}-\frac{9}{4} \frac{1}{(N+1)^2}
+\frac{1}{4}\frac{1}{N^3}-\frac{1}{4}\frac{1}{(N+1)^3}
\bigg] {A(m^2)}^2\,+\, {\cal O}({A(m^2)}^3)
\right\} \; .
\end{eqnarray}
The order $\alpha_s$ result is known. It was extracted by Mele and
Nason~\cite{MN}, starting from the single inclusive cross section
for heavy-quark production in $e^+e^-$ annihilation and
subtracting the coefficient function for the massless case. More
recently, this term has been computed in a process-independent way
in~\cite{CC}. Our calculation extends it to all orders in the
large-$\beta_0$ limit.

In addition to the resummation of large perturbative corrections,
the all-order  large-$\beta_0$ result can be used to extract some
non-perturbative information on the fragmentation process. As we
saw, in the large-$\beta_0$ limit renormalons appear always as
simple poles, and are exclusively associated with the integration
over $x$. It should be noted that the subtracted
term~(Eq.~(\ref{B_E})) has no infrared renormalons in
factorization schemes such as ${\overline{\rm MS}}$.

In~(\ref{B_D}) the first singularity is located at $u=1/2$ and it
corresponds to ${\cal O}(\Lambda/m)$ corrections. The second
singularity is at $u=3/2$ (${\cal O}(\Lambda^3/m^3)$ corrections) and
higher singularities appear at all integers and half integers on the
positive Borel axis.  It is particularly interesting to note the
absence of the renormalon at $u=1$. As explained above this structure
can be traced back to the relation between the single and the double
pole terms in $2p_1k+k^2$ in the squared matrix element
(\ref{M2_dominant}).

Both the presence on the renormalon singularity at $u=1/2$ and the
absence of the one at $u=1$ are consistent with the findings of Nason
and Webber~\cite{Webber_Nason}, which considered the large-$N$ limit of
the fragmentation function in $e^+e^-$ annihilation. We confirm these
conclusions  in a more general, process independent context (and for a
generic $N$), and derive an expression for the Borel function allowing
to combine resummation with parametrization of power corrections. It is
not known, and certainly deserves further investigation, whether a
renormalon at $u=1$ does appear in the full theory.

\subsubsection{The large-$x$ region and the Sudakov exponent in perturbation
theory and beyond}

At large $x$ multiple emission plays an important r{\^o}le. The SDG result can
be  readily used to derive the exponentiation kernel in the approximation of
independent emission.
Extracting the log-enhanced terms from~(\ref{B_D}) and~(\ref{B_E}) one gets
\ba
\label{D_N_MSbar}
\!\!\!\!\!\left.\ln \tilde{D}(N,m^2)\right\vert_{\DGE}=
\frac{ C_F}{\beta_0}\,\int_0^{\infty}\,du
\left[-\frac{B_{\tilde{D}}^{\DGE}(u,N)}{u}\,
\,+\,\frac{\,B_{\cal A}(u)}{u}\,\ln N \right]
\, \,T(u)\,\left(m^2/{\Lambda}^{2}\right)^{-u}
\ea
where $\mu$ was set equal to $m$,
\begin{eqnarray}
\label{BD}
B_{\tilde{D}}^{\DGE}(u,N)=-\,{\rm e}^{cu}\,
(1-u)\,\Gamma(-2u)\,\left[N^{2u}-1\right]\,+\,{\cal O}(1/\beta_0),
\end{eqnarray}
and $B_{\cal A}(u)=1+c\frac{u}{1!}+\ldots$ is the Borel transform
of the cusp
 anomalous dimension~(\ref{G_large_Nf}).

Note that we write here a perturbative expansion for $\ln \tilde{D}(N,m^2)$.
This is not just a formal manipulation which holds to order $\alpha_s$, but
rather a statement about the structure of higher-order terms.
The logarithmically enhanced terms,  as opposed to the expressions in
(\ref{N_ren_sum}), (\ref{N_space_result}) and (\ref{large_beta0_exp}),
{\em  exponentiate}: they can be written to all orders as
\[
\tilde{D}(N,m^2)=\exp\left\{\sum_{n=0}^{\infty}\sum_{k=1}^{n+1}
d_{n,k}\,
\alpha_s^n \ln^kN\right\}.
\]
Although~(\ref{D_N_MSbar}) was derived strictly in the
large~$\beta_0$ limit, i.e. leading-log (LL) accuracy, it can be
generalized in a straightforward way to comply with NLL accuracy
in the full theory~\cite{CMW,Gardi:2001ny,DGE} by using the
two-loop running coupling~(\ref{A_B}) and replacing $c$ of
Eq.~(\ref{c}) in the exponent ${\rm e}^{cu}$ and in $B_{\cal
A}(u)$ by \be c\,\longrightarrow\,
a_2=5/3+(1/3-\pi^2/12)C_A/\beta_0. \label{a_2} \ee It should be
noted, though, that the functional form of the Borel transform is
not known beyond the large-$\beta_0$ limit, so the replacement of
$c$ by $a_2$ in~(\ref{BD}) is not proven. As discussed in a
similar context in~\cite{Gardi:2002xm} the same logarithmic
accuracy can be achieved in different ways, e.g. by replacing the
factor ${\rm e}^{(5/3)u}$ in~(\ref{BD}) by ${\rm
e}^{(5/3)u}(1+u(a_2-5/3))$. We will assume here the simple
replacement (\ref{a_2}).

The renormalon structure of the Sudakov exponent, is similar to that of
the full SDG result~(\ref{B_D}): the leading renormalon is at $u=1/2$
and subleading ones appear at all integer and half integers $u\geq
3/2$. In accordance with the general property of the fragmentation
function (Sec.~2) these corrections show up in~(\ref{D_N_MSbar}) on
the scale $m/N$. The appearance of these powers corrections in the
exponent leads to factorization, in moment space, of the
non-perturbative contribution to the fragmentation function. Moreover,
it also implies that these corrections exponentiate together with the
perturbative logs.

Based on the result of Sec.~2 we can further deduce that in the
full theory, as in (\ref{D_N_MSbar}), renormalons in the Sudakov
exponent appear as single poles, involving no anomalous dimension.
In case on non-vanishing anomalous dimensions~$\gamma_i$, power
corrections are modified by logarithms, $(\mu N/m)^i (\ln
\mu/m)^{\gamma_i/(2 \beta_0)}$ where $\mu$ is a cutoff. This
behavior contradicts Eq.~(\ref{large_N_JR}) and therefore the
anomalous dimensions must vanish.

We conclude that non-perturbative corrections to the heavy-quark
fragmentation process appear as
\be
\tilde{D}(N,m^2)=\tilde{D}_{\PT}(N,m^2)\,\tilde{D}_{\NP}(N\Lambda/m),
\label{psf}
\ee
where $\tilde{D}_{\PT}(N,m^2)$ is obtained by exponentiating the
(regularized) renormalons sum~(\ref{D_N_MSbar}) and
\be
\tilde{D}_{\NP}(N\Lambda/m)=\exp\left\{-\omega_1\frac{C_F}{\beta_0}
\frac{N\Lambda}{m}-\omega_3\frac{C_F}{\beta_0}
\left(\frac{N\Lambda}{m}\right)^3-\omega_4\frac{C_F}{\beta_0}
\left(\frac{N\Lambda}{m}\right)^4+\cdots
\right\},
\label{D_NP}
\ee
which is characterized by the absence of the second power of $(N\Lambda
/m)$. Here $\omega_n$ are left as free parameters. These parameters are
defined  only within a given regularization prescription for the
renormalons,  corresponding to a definite separation between the
perturbative and non-perturbative components in~$\tilde{D}(N,m^2)$. A
physically meaningful separation would be the one based on a momentum
cutoff, or, alternatively, a principle-value regularization of the
Borel sum, which can be readily related to a
cutoff~\cite{Gardi:1999dq}. Having performed the renormalon sum
in~$\tilde{D}_{\PT}(N,m^2)$, the parameters in
$\tilde{D}_{\NP}(N\Lambda/m)$ can be fixed by comparison with
experimental data.

Going over to $x$ space, the product in (\ref{psf}) becomes a
convolution with a {\em shape function} depending on $m(1-x)$. As
discussed in the context of event-shape distributions in $e^+e^-$
annihilation~\cite{DW,KS,Korchemsky:2000kp,Gardi:2001ny}, this
convolution generates a {\em shift} of the perturbative distribution
proportional to~$\omega_1$ and a deformation of its shape by $\omega_3$
and on.

It should be stressed that both the resummation
formula~(\ref{D_N_MSbar}) and the   non-perturbative corrections
in~(\ref{D_NP}) are expected to dominate only for $N\gg 1$. The first
few moments are surely affected by perturbative and non-perturbative
contributions on the scale $m$ (as opposed to $m/N$), which do not necessarily exponentiate and were neglected here.  The phenomenological implications
of~(\ref{D_N_MSbar}) and~(\ref{D_NP}) will be discussed in more detail
in the context of $e^+e^-$ annihilation.

\section{Heavy meson production in $e^+e^-$ annihilation}

\subsection{Calculation of the Sudakov exponent}

Let us consider now the specific process of heavy-flavor
production in $e^+e^-$ annihilation. To be concrete, let us
concentrate on the vector current contribution to the process
$e^+e^-\longrightarrow Q\bar{Q}g$ where the momentum of the
virtual photon is $q$ (the centre-of-mass energy is $\sqrt{q^2}$),
and the quarks have momenta $p_1$ and $p_2$, energy fractions
$x\equiv 2p_1q/q^2$ and $\bar{x}\equiv 2p_2q/q^2$ and masses
$p_1^2=p_2^2=m^2\equiv \rho\,q^2/4$.

As in~\cite{Webber_Nason} we begin with the exact
matrix element with an off-shell
gluon, $k^2  \equiv \epsilon q^2$. The result for the cross section is,
\begin{eqnarray}
\label{mat_ele}
\frac{1}{\sigma}\frac{d\sigma}{dx\, d{\bar{x}}}&=&
\frac{C_F\alpha_s}{2\pi}\,\frac{1}{\sqrt{1-\rho}}
\left[\frac{(x+\epsilon+\rho/2)^2+({\bar{x}}+\epsilon+\rho/2)^2-2\rho
(1+\epsilon+\rho/2)}{(1+\rho/2)(1-x)(1-\bar{x})}\right. \nonumber \\
&&\left. -\frac{\epsilon+\rho/2}{(1-x)^2}
-\frac{\epsilon+\rho/2}{(1-\bar{x})^2}\right].
\end{eqnarray}
It is straightforward to verify that when taking the
quasi-collinear limit this expression reduces to the approximate
one we calculated in the previous section in a process-independent
way. Note first that the propagator which becomes singular in this
limit is inversely proportional to
$2p_1k+k^2=q^2-2p_2q=q^2(1-\bar{x})$, where we used momentum
conservation: $k=q-p_1-p_2$. As follows from Eq.~(\ref{prop}), the
quasi-collinear limit is the one in which  $1-\bar{x}$, $\epsilon$
and $\rho$ (and $k_{\perp}^2/q^2$) all become small but the ratios
between them are fixed. In this limit the leading contribution
to~(\ref{mat_ele}) takes the form:
\begin{eqnarray}
\label{app_mat_ele}
\frac{1}{\sigma}\frac{d\sigma}{dx\, d{\bar{x}}}\simeq
\frac{C_F\alpha_s }{2\pi}\,
\left[\frac{1+x^2}{1-x}\frac1{(1-\bar{x})}-\frac{\rho/2+\epsilon}
{(1-\bar{x})^2}\right],
\end{eqnarray}
which fully agrees\footnote{To verify the consistency of the two expressions
one should write $p_2$ in terms of the Sudakov parameters we used in the
previous section:
\[
p_2^\nu=bp^\nu+a(2pn)n^\nu,
\]
\be
a=\frac{1-b^2}{b}\frac{m^2}{(2pn)^2}, \quad \quad
b=\frac{\bar{x}-\rho/2-\sqrt{\bar{x}^2-\rho}}{2(1-\bar{x})+\rho/2},
\ee
where $a$ and $b$ were fixed by the conditions $p_2^2=m^2$ and
$(k+p_1+p_2)^2=q^2$.
Next, evaluating ${\rm Tr}(\psl{\cal M}_0{\bar {\cal M}_0})=8pp_2=
4m^2(1+b^2)/b$,
one finds that in the limit considered $a\simeq {q^2}/{(2pn)^2}$,
${\rm Tr}(\psl{\cal M}_0{\bar {\cal M}_0})=4q^2$ and $x\simeq z$, where
terms linear in one of the small parameters were neglected. Finally,
making these substitutions in (\ref{M2_dominant}), writing the phase-space
measure as $d\phi=q^2/(16\pi^2)\,dz\,d\bar{x}$ and dividing by the flux
factor $1/(4q^2)$, one find that the approximate result of the previous
section indeed coincides with~(\ref{app_mat_ele}).} with (\ref{M2_dominant}).

Owing to the non-vanishing quark mass and to its
inclusive nature, the differential cross section
$d\sigma/dx$ is an infrared and collinear safe quantity. It can
therefore be calculated in perturbation theory to any order, with no
need to perform factorization.

Here we perform such a calculation having in mind the experimentally
interesting scenario with $\rho\ll 1$ as well as $1-x\ll 1$ (or $N\gg
1$). These conditions are easily met for charm and bottom production at
LEP1 energies, i.e. $q \simeq 91~{\rm GeV}$.
In this limit logarithmically enhanced terms of either $\rho$ or
$N$ need to be resummed. At the same time, in order to address the
issue of power corrections, we must perform renormalon resummation.
This calls for DGE.

In this section we  perform the calculation
by integrating the full off-shell gluon matrix
element~(\ref{mat_ele}) over the exact phase space.  This calculation
follows the steps of the first derivation of the DGE result for the
thrust distribution~\cite{Gardi:2001ny}, based on the characteristic
function of~\cite{moments}: It proceeds by identifying
log-enhanced terms and deriving a Borel representation which generates
these terms to all orders in the large-$\beta_0$ limit. Finally, this
Borel sum is used as the exponentiation kernel.

\begin{figure}[htb]
\begin{center}
\epsfig{file=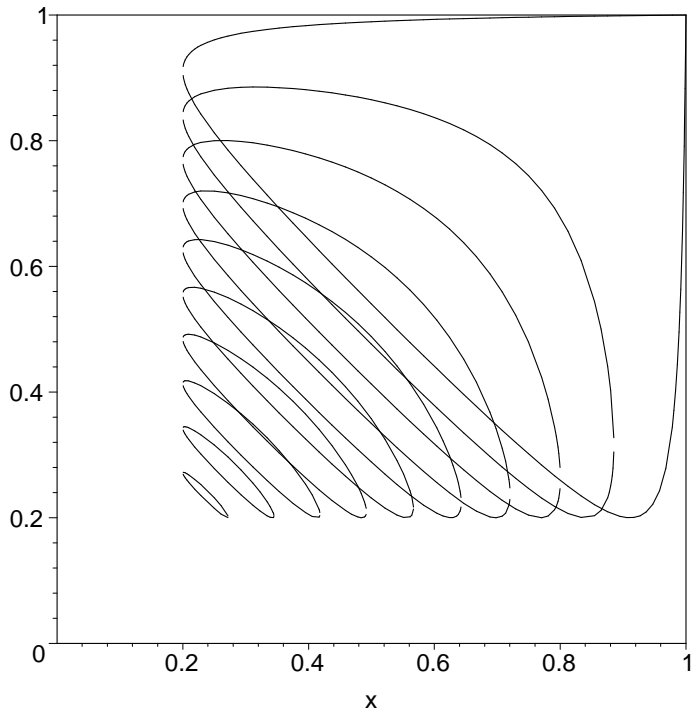,width=9cm}
 \caption{$QQg$ phase-space boundaries in the $x$--$\bar{x}$ plane, with
 varying gluon virtualities \hbox{$\epsilon=(1-\sqrt{\rho})^2\,i/10$},
 for $i=0$ through $9$, with $\rho=0.04$.}
\label{phase-space}
\end{center}
\end{figure}

First we integrate~(\ref{mat_ele}) over $\bar{x}$, getting a characteristic
function.
The limits of phase space (see~\cite{Webber_Nason}) are:
\begin{eqnarray}
\bar{x}_{\max /\min}&=&\frac{(2-x)(1-x-\epsilon+\rho/2)\pm\Xi}{2(1-x)+\rho/2}
\nonumber \\
\Xi&=&\sqrt{x^2-\rho}\sqrt{(1-x-\epsilon)^2-\epsilon\rho}.
\end{eqnarray}
As Fig.~\ref{phase-space} shows, for a given $\rho$ the phase-space shrinks as
$\epsilon$ increases. Note that even with a vanishing gluon virtuality the
phase-space boundary does not reach the line $\bar{x}=1$ (with the exception
of a point at $x=1$), where the matrix element is singular. Consequently the
differential cross section $d\sigma/dx$ is infrared and collinear safe.
The resulting characteristic function~is:
\be
\label{char_func}
\frac{1}{\sigma}\frac{d\sigma}{dx}(\epsilon,x,\rho)=
\frac{C_F\alpha_s}{\pi}\left[{\cal F}_1+{\cal F}_2+{\cal F}_3\right]
\ee
with
\begin{eqnarray}
{\cal F}_1 &=& \frac{- (\rho+2 \epsilon)}{\rho \xi^2+4 \epsilon-4 \epsilon
\xi+4 \epsilon^2-4 \epsilon \rho} \,\Xi \,\nonumber \\
{\cal F}_2 &=& \frac{4-4 \xi+2 \xi^2+8 \epsilon-4 \epsilon \xi-2 \rho \xi+4
\epsilon^2-\rho^2}{2\xi(2+\rho)}\,\ln\left(\frac{2 \xi+2 \epsilon-2 \xi^2+2
\epsilon \xi-\rho \xi+2\, \Xi}{{2 \xi+2 \epsilon-2 \xi^2+2 \epsilon \xi-\rho
\xi-2 \,\Xi}}\right) \nonumber \\
{\cal F}_3 &=& \left(\frac{24\epsilon}{\xi(2+\rho)\rho^2}-\frac{ \rho+2
\epsilon}{\rho \xi^2}-\frac{2 (\rho^2+48 \epsilon)}{(2+\rho)\rho^2(4 \xi+\rho)}+
\frac{2 (-4+\rho) (-\rho+4 \epsilon)}{(2+\rho)\rho(4 \xi+\rho)^2}\right) \,\Xi,
\end{eqnarray}
where $\xi\equiv 1-x$. These three terms are distinguished by the singularity
of the corresponding term in the matrix element at $\bar{x}\longrightarrow 1$:
${\cal F}_1$ originates in a double pole, ${\cal F}_2$ in a single pole and
${\cal F}_3$ in non-singular terms.

Next we perform the integral over the gluon virtuality $\epsilon$
in the range $0<\epsilon<\epsilon_{\max}$, where the upper
integration limit, \be \epsilon_{\max}
=\frac14\,\left[\sqrt{\rho}-\sqrt{\rho+4(1-x)}\right]^2 \ee is the
point where $\Xi=0$ and $\bar{x}_{\min}=\bar{x}_{\max}$. The Borel
function is obtained by computing: \be B(u,x,\rho)=-\frac{\sin \pi
u}{\pi u}\,{\rm e}^{cu}\,\,I(u,x,\rho),\quad\quad
I(u,x,\rho)\equiv u\,\int_0^{\epsilon_{\max}}d\epsilon\,
\epsilon^{-1-u}\, {\cal F}(\epsilon,x,\rho). \ee The exact
integration is complicated. The full result can only be written in
terms of hypergeometric functions. However, in the following we
will not need the full result, as we are interested in the
logarithmically enhanced terms alone. Assuming the hierarchy
$\rho\ll \xi\ll 1$ the leading terms in the Borel
 function are rather simple. The integrals corresponding to the three terms in
 the characteristic functions are the following:
\begin{eqnarray}
I_1 &\simeq& \,\frac{\pi u}{\sin \pi u}\, \xi^{-1-2u} (\rho/4)^{-u},\nonumber \\
I_2 &\simeq& \frac{1}{u}\,\xi^{-1-u}\,-\,\frac{\pi }{\sin \pi u}\, \xi^{-1-2u}
(\rho/4)^{-u}                 \nonumber \\
I_3 &\simeq& \frac12 \xi^{-1-u}\,\left[\frac{1}{1-u}+\frac{1}{2-u}\right]
\end{eqnarray}
where we neglected terms suppressed by powers of either $\xi$ or $\rho$.
The Borel function which generates all the log-enhanced terms in the large
$\beta_0$ limit is therefore given by
\begin{eqnarray}
\label{B_sigma_y}
B(u,x=1-\xi,\rho)&=&\frac{{\rm e}^{cu}}{\xi} \,\bigg\{ \frac{1}{u}\,
(1-u)\,\left(\rho/4\right)^{-u}\, \xi^{-2u}
\nonumber \\
&&\hspace*{40pt}-\, \frac{\sin \pi u}{2\pi u}\left(\frac{2}{u}+\frac{1}{1-u}+
\frac{1}{2-u}\right)\,
\, \xi^{-u}\bigg\}.
\end{eqnarray}

Finally we perform the $x$ integration and incorporate virtual
corrections through the subtraction of the result at $N=1$. In
spite of the fact that (\ref{B_sigma_y}) was derived assuming the
hierarchy $\rho\ll \xi\ll 1$, it is sufficient to get the exact
result for the Sudakov exponent, since the region of
$\xi\lsim\rho$ yields contributions which are suppressed by powers
of~$\rho$. The resulting Borel representation of the Sudakov
exponent is \be \left.\ln
\tilde{\sigma}(N,q^2,m^2)\right\vert_{\DGE}= \frac{
C_F}{\beta_0}\,\int_0^{\infty}\,du B_\sigma(u,N,m^2/q^2)\,
\,T(u)\, \exp(-u\ln q^2/{\Lambda}^{2}) \label{sigma_N} \ee with
\begin{eqnarray}
\label{B_sigma}
B_\sigma(u,N,m^2/q^2)&=&{\rm e}^{cu}\, \bigg\{ \frac{1}{u}\,
(1-u)\,\Gamma(-2u)\,\left[N^{2u}-1\right]\,\left({m^2}/{q^2}\right)^{-u}
\nonumber  \\ &&\hspace*{30pt} -\frac{\sin \pi u}{2\pi u}\left(\frac{2}{u}+
\frac{1}{1-u}+\frac{1}{2-u}\right)\,
\Gamma(-u)\,\left[N^u-1\right]\bigg\}.
\end{eqnarray}

\subsection{Factorization}

Our result~(\ref{sigma_N}) for the Sudakov exponent of the
$e^+e^-\longrightarrow Q+X $ differential cross section involves
two external scales, $q^2$ and $m^2$. Therefore, in addition to
Sudakov logs $L\equiv \ln N$, it contains collinear logs $l\equiv
\ln q^2/m^2$. The latter can be seen a reflection of evolution
from the scale $m^2$ at which the fragmentation process of the
heavy quark takes place (Sec.~2 and 3) to the
$e^+e^-$~centre-of-mass energy squared  $q^2$ at which the hard
interaction takes place. The contributions to the cross section
from these two subprocesses  can be written separately by
introducing some factorization procedure. Since Sudakov logs
emerge from both subprocesses, this separation will be useful to
distinguish between them.

Factorization takes the form of a product in moment space, so the Sudakov
exponent can be written as a {\em sum} of the following three exponents,
\begin{eqnarray}
\label{sigma_N_fact}
\left.\ln \tilde{\sigma}(N,q^2,m^2)\right\vert_{\DGE}=
\,\left.\ln \tilde{J}(N,q^2)\right\vert_{\DGE}\,
+\,\left.\ln \tilde{E}(N,q^2,m^2)\right\vert_{\DGE} \,
+\left.\ln \tilde{D}(N,m^2)\right\vert_{\DGE}\,
\end{eqnarray}
where $\ln \tilde{D}(N,m^2)$ corresponds to Eq.~(\ref{D_N_MSbar}),
i.e. the (process independent) heavy-quark fragmentation depending
on the scale $m^2$, $\ln \tilde{J}(N,q^2)$ corresponds to the
$e^+e^-$~coefficient function depending on the scale $q^2$, and
$\ln \tilde{E}(N,q^2,m^2)$ represents the evolution between these
two scales and thus it depends on both.

Performing factorization, the Sudakov exponent corresponding to
the fragmentation subprocess is~(\ref{D_N_MSbar}), the evolution
is given by \ba \label{E_N} \!\! \left.\ln
\tilde{E}(N,q^2,m^2)\right\vert_{\DGE}= \frac{
C_F}{\beta_0}\,\int_0^{\infty}du\, \,T(u)\,
\left(q^2/{\Lambda}^{2}\right)^{-u}\,\frac{B_{\cal A}(u)}{u}\, \ln
N \,\left\{1\,-\,\left({m^2}/{q^2}\right)^{-u}\right\}\, \ea and
the Sudakov exponent of the $e^+e^-$ coefficient function is \ba
\label{J_N} \left.\ln \tilde{J}(N,q^2)\right\vert_{\DGE}= &&
-\frac{ C_F}{\beta_0}\,\int_0^{\infty}\,du
\,T(u)\,\left(q^2/{\Lambda}^{2} \right)^{-u} \,\\ \nonumber
&&\bigg\{{\rm e}^{cu}\, \frac{\sin \pi u}{2\pi u}
\left(\frac{2}{u}+\frac{1}{1-u}+\frac{1}{2-u}\right)\,
\Gamma(-u)\,\left[N^u-1\right] \,+\,\frac{B_{\cal A}(u)}{u}\,\ln
N\bigg\}. \, \ea The essential ingredient in obtaining a
factorized formula is that each of the three exponents
corresponding to the separate subprocesses has a well-defined
perturbative expansion: there are no $1/u$ singularities in these
Borel functions. Of course, there is some arbitrariness in this
procedure as far as higher-order terms are concerned. Here we
chose the natural factorization scales as the external scales,
$m^2$ and $q^2$. The subtracted term $B_{\cal A}(u)=1+a_2u+\ldots$
depends on the factorization scheme. In ${\overline {\rm MS}}$,
the coefficient $a_2$ is given by~(\ref{a_2}).     Owing to the
fact that the expressions above correspond to all-order
resummation, factorization-scale and scheme dependent terms cancel
out completely in the sum~(\ref{sigma_N_fact}). Of course, if any
of the three exponents is replaced by some fixed order or a
fixed-logarithmic accuracy approximation, some factorization-scale
and scheme dependence will appear in the final result for the
cross section.

Note that while~$\tilde{J}(N,q^2)$ and~$\tilde{D}(N,m^2)$ contain
Sudakov double logs (upon expanding the exponent, the highest power of
$\ln N$ is twice that of the coupling at each order), the evolution
factor $\tilde{E}(N,q^2,m^2)$  contains at most $\ln N$ to the same power
as the coupling. This is so because $\ln \tilde{E}(N,q^2,m^2)$ has {\em
just one power} of $\ln N$ to any
order in $\alpha_s$~\cite{Sterman:1986aj,Korchemsky:1988si,KM}.

Having performed factorization we can compare the results obtained for
the two subprocesses~(\ref{D_N_MSbar}) and~(\ref{J_N}) to the
corresponding process-independent calculations. The DGE Sudakov
exponent for the heavy-quark fragmentation subprocess  has been
identified with the result obtained in the process-independent
calculation~(\ref{D_N_MSbar}).  It follows that the only scale which
plays a r{\^o}le in the fragmentation process $\tilde{D}(N,m^2)$ at large
$N$ is $m/N$. As shown in Sec.~2, this is a general property of the
fragmentation function which holds beyond the perturbative level.

Similarly, the DGE Sudakov exponent for the $e^+e^-$ coefficient
function~(\ref{J_N})  can be identified as the one appearing in a
massless quark fragmentation process which was calculated
in~\cite{DGE}. The same all-order result was
obtained~\cite{DGE,DIS,Gardi:2002xm} for the $F_2$ structure function
at large $N$. As discussed in~\cite{DGE,DIS},  $\left. \ln
\tilde{J}(N,q^2)\right\vert_{\DGE}$ is associated with radiation from
the undetected quark (or the outgoing quark in a deep-inelastic process
at large $x_{\rm Bj}$) and the formation of a jet
(this explains the notation $\tilde{J}$) with invariant mass
$q^2/N$, which recoils against the detected quark. This identification
means, in particular, that $\left. \ln
\tilde{J}(N,q^2)\right\vert_{\DGE}$ is not specific to the particular
process in which two heavy quarks are produced: it would be the same if
the recoiling jet would have been formed around a light quark. The
internal structure of the jet is not resolved as the relevant scale is
its invariant mass.

\begin{figure}[htb]
\begin{center}
\epsfig{file=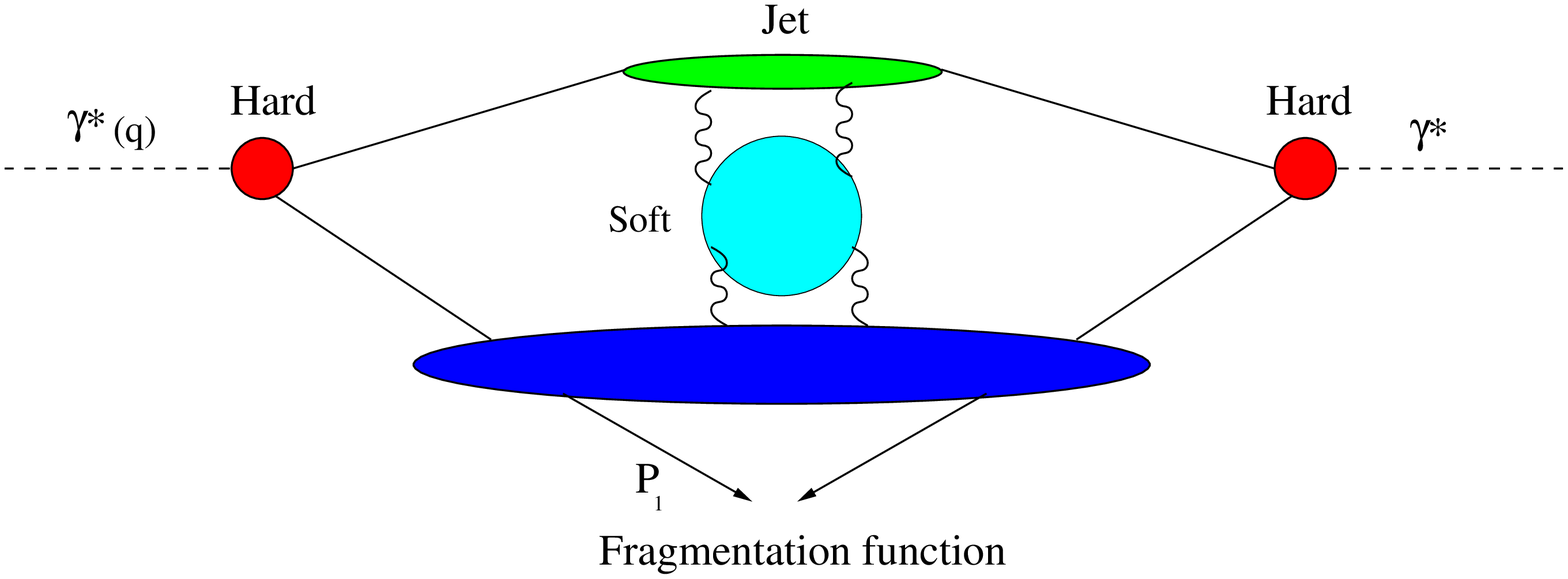,width=15cm}
\caption{Factorization of the process $\gamma^*\longrightarrow Q+X$ at
large $N$.}
\label{Fact}
\end{center}
\end{figure}

The fully factorized process of heavy-quark production in $e^+e^-$
annihilation is shown in Fig.~\ref{Fact}. The hard blob represents
radiative corrections with virtualities of order of the centre-of-mass
energy squared~$q^2$. The detected heavy meson ($p_1$) emerges from the lower
jet in the figure. Although a priori two-jet production is symmetric,
the fact that one measures a single particle  inclusive cross section
breaks the symmetry: the measurement is sensitive to different features
of the two jets.  The fragmentation blob represents $\tilde{D}(N,m^2)$.
It accounts for soft and collinear radiation on the scale $m/N$. The
recoiling jet on the upper part of the figure represents
$\tilde{J}(N,q^2)$, accounting for soft and collinear radiation of the
scale $q^2/N$. The soft blob is associated with the evolution factor
$\tilde{E}(N,q^2,m^2)$. It accounts for gluons whose momentum
components are all small. These gluons interact with eikonalized quarks:
They are insensitive to the invariant mass or the structure of either of
the two jets.

Alongside the dominant perturbative corrections which are
summarized by~(\ref{sigma_N_fact}), the large-$N$ limit singles
out certain non-perturbative corrections. Such corrections appear
in two of the subprocesses described above: in $\tilde{D}(N,m^2)$
they appear on the scale $m/N$ and in $\tilde{J}(N,q^2)$ on the
scale $q^2/N$. The evolution factor $\tilde{E}(N,q^2,m^2)$, on the
other hand, has no power corrections. The presence of power
corrections, as well as their parametric form, can be deduced from
the renormalon singularities in the corresponding Borel sums. In
both $\tilde{D}(N,m^2)$ and $\tilde{J}(N,q^2)$, the dominant
corrections at large $N$ appear in the Sudakov exponents,
Eq.~(\ref{D_N_MSbar}) and (\ref{J_N}), so these power corrections
{\em exponentiate} together with the perturbative logs. The
renormalon singularities of $\left. \ln
\tilde{J}(N,q^2)\right\vert_{\DGE}$ were analyzed
in~\cite{DGE,DIS,Gardi:2002xm}. They lead to $\Lambda^2 N/q^2$ and
$\Lambda^4 N^2/q^4$ corrections at the exponent, as summarized by
Eq.~(50) in~\cite{DGE}. As discussed in the previous section the
renormalon singularities of the heavy-quark fragmentation exponent
$\left. \ln \tilde{D}(N,m^2)\right\vert_{\DGE}$ appear at $u=1/2$
and at any integer and half integer such that $u\geq 3/2$. There
is no renormalon pole at $u=1$ owing to the factor $(1-u)$
in~(\ref{BD}).

\section{Implications for phenomenology}
\label{sec:nlldge}

In the previous sections we studied the problem of heavy-quark
fragmentation, concentrating on the large-$N$ limit. We began by
identifying the formal limit in which simplification occurs, namely
where $N$ and $m$ get large simultaneously with the ratio $m/N$ fixed.
We then performed a  process-independent  perturbative calculation by
DGE, facilitating the resummation of Sudakov  logs as well as that of
running-coupling effects. The combined treatment of  Sudakov  logs and
renormalons sets the basis for a systematic parametrization of
power-suppressed corrections on the scale $m/N$, which exponentiate
together with  the perturbative logs. Then, specializing to the case of
$e^+e^-$ annihilation, we demonstrated  how this result for the Sudakov
exponent of the fragmentation  function emerges from the
process-specific renormalon calculation upon factorization. The
Sudakov-resummed coefficient  function in the $e^+e^-$ case has been
identified with the familiar  jet function which plays an important
r{\^o}le in light-quark  fragmentation and in deep inelastic structure
functions at large~$x$.

The purpose of this section is to study the implications of these new
results on phenomenology. While there are many possible applications,
we concentrate here on the one where most precise data is available,
namely bottom production is~$e^+e^-$ annihilation at LEP1. In this study we
work directly with data in moment space. This is a natural choice
from a theoretical perspective, but, as we will see, not an easy
one from the experimental point of view: the presence of strong correlations
between different moments requires a very careful treatment of the errors.

In our phenomenological analysis we will
concentrate on the large-$x$ region. This
region is inaccessible with a perturbative approach
without Sudakov resummation.
It is still very problematic when standard NLL resummation is
applied~\cite{CC}, as the perturbative spectrum is significantly
more peaked than the data and it becomes negative near $x=1$.
In such circumstances, matching the perturbative result with
the non-perturbative contribution becomes awkward.

This is not intended to be an exhaustive phenomenological analysis of
the heavy-quark fragmentation function.  We concentrate on improving the
resummation at large~$x$ and the parametrization of power corrections
on the scale $m/N$, and we simplify other aspects. We do not take into
account power-suppressed corrections in $m^2/q^2$, nor do we include
${\cal O}(\alpha_s^2)$ terms which are free of~$\ln m^2/q^2$ and $\ln
N$ enhancement. Such effects have been considered
elsewhere~\cite{Nason:1999zj},
and found to have a limited impact.

In Sec.~\ref{sec:pt} we study the implication of our approach
at the perturbative level, comparing it to NLL Sudakov resummation. We
also detail there the prescription we use for matching the resummed
result with the fixed-order calculation and the way we deal with
infrared renormalon ambiguities. In Sec.~\ref{sec:npfit} we address
the non-perturbative contribution to the fragmentation function. We
first discuss the data and the way we use them, and then present  results
of various fits where $\alpha_s$ and the non-perturbative parameters
are determined by the data.

\subsection{The perturbative result}
\label{sec:pt}
\subsubsection{Factorization, matching and regularization of renormalon
singularities}

Making full use of the concept of factorization as outlined
in~\cite{CC}, we write the normalized Mellin moments of the
$e^+e^-$ differential cross section $(1/\sigma)(d\sigma/dx_E)$ as
a product \beq \tilde{\sigma}(N,q^2,m^2) \,=\,
\tilde{C}(N,q^2;\mu_F^2) \, \tilde{E}(N,\mu_F^2,\mu_{0F}^2)
\,\tilde{D}(N,m^2;\mu_{0F}^2)\; , \label{eq:eefull} \eeq where
$\tilde{D}(N,m^2;\mu_{0F}^2)$ is the process-independent
heavy-quark fragmentation
function\footnote{$\tilde{D}(N,m^2;\mu_{0F}^2)$ is sometimes
called the ``initial condition'' for the evolution. This term was
therefore given the label ``ini'' in~\cite{CC}.} defined in
Eqs.~(\ref{D_def}) and~(\ref{D_mom}), $\tilde{C}(N,q^2,\mu_F^2)$
is an $e^+e^-$ massless coefficient function, and
$\tilde{E}(N,\mu_F^2,\mu_{0F}^2)$ is an $\overline {\rm MS}$
Altarelli-Parisi evolution factor, given in Eq.~(43) of~\cite{CC},
which is associated with the ultraviolet singularity of
$\tilde{D}(N,m^2;\mu_{0F}^2)$ (see e.g.~(\ref{N_space_result}))
and the collinear singularity of $\tilde{C}(N,q^2;\mu_F^2)$. We
will choose the natural factorization scales: $\mu_F^2=q^2$ and
\mbox{$\mu_{0F}^2 = m^2$}. Varying these scales was
shown~\cite{CC} to have a small effect, provided that resummation
is performed in each of these functions.

In contrast with the evolution factor, the fragmentation function
$\tilde{D}(N,m^2)$ and the coefficient function $\tilde{C}(N,q^2)$
contain double logs of $N$ and  infrared renormalons.  If
$q^2/N\ll\Lambda^2$ the coefficient function can be safely
calculated using Sudakov resummation to NLL accuracy, as done
in~\cite{CC}. The same holds for  the fragmentation function if
$m/N\ll\Lambda$. The latter, however, holds for the first few
moments at best. Therefore, here the effects of renormalon
resummation and the corresponding power corrections are expected
to be important for phenomenological applications.

The Sudakov exponents corresponding to the fragmentation function and
the coefficient function are given to all orders in the large-$\beta_0$
limit in Eqs.~(\ref{D_N_MSbar})
and~(\ref{J_N}), respectively. These exponents can be employed in an
expanded form, e.g. truncated at NLL order,
\beq
\left.\ln \tilde{D}(N,m^2)\right\vert_{\NLL} =
g^{(1)}_{\rm ini} \ln N  + g^{(2)}_{\rm ini} \; ,
\eeq
where $g^{(1)}_{\rm ini}$ and $g^{(2)}_{\rm ini}$ are given in
eqs. (74) and (75) of Ref.~\cite{CC}, or evaluated
in full, thus providing some resummation of subleading logs.
In either case, we match\footnote{Note that the matching procedure
used in~\cite{CC} is somewhat
different. The numerical differences are, however, negligible.}
the exponents to the fixed-order results at
${\cal O}(\alpha_s)$ by the so-called `log-$R$ matching' procedure:
\beq
\ln \tilde{D}(N,m^2) = \left.\ln \tilde{D}(N,m^2)\right\vert_{\DGE} +
\left.\ln \tilde{D}(N,m^2)\right\vert_{\rm fixed\, order} -
\left.\ln \tilde{D}(N,m^2)\right\vert_{\rm overlap} \;,
\label{eq:matching}
\eeq
where $\left.\ln \tilde{D}(N,m^2)\right\vert_{\DGE}$
is given by~(\ref{D_N_MSbar}),
$\left.\ln\tilde{D}(N,m^2)\right\vert_{\rm fixed\, order}$
is the expansion of the logarithm of the
 full result\footnote{
 This result was first obtained in
\cite{MN}; see Eq.~(A.13) there. It can also be read off
Eq.~(\ref{large_beta0_exp}) of the present paper.} for
$\tilde{D}(N,m^2)$ truncated at ${\cal O}(\alpha_s)$ and \beq
\left.\ln \tilde{D}(N,m^2)\right\vert_{\rm overlap} =
\frac{\alpha_s(m^2)}{\pi} \, C_F\; \left[ - \ln^2N + (1 -
2\gamma_E  ) \ln N \right] \eeq is subtracted to account for the
terms which are double counted when adding the two previous
contributions. Note that we suppress here the explicit dependence
on the renormalization and the factorization scales (see Eq.~(45)
in~\cite{CC}), which are both set equal to~$m$. The matching of
the coefficient function~$\tilde{C}(N,q^2)$ is done in the same
way:
 \beq
\ln \tilde{C}(N,q^2) = \left.\ln
\tilde{J}(N,q^2)\right\vert_{\DGE} + \left.\ln
\tilde{C}(N,q^2)\right\vert_{\rm fixed\, order} - \left.\ln
\tilde{J}(N,q^2)\right\vert_{\rm overlap} \;,
\label{eq:matching-cf} \eeq where $\left.\ln
\tilde{J}(N,q^2)\right\vert_{\DGE}$ is given in (\ref{J_N}) and
$\tilde{C}(N,q^2;\mu_F^2)$, which includes contributions to the
hard subprocess as well as the jet (see Fig.~\ref{Fact}) is given 
at ${\cal O}(\alpha_s)$ in Eq.~(A.12) of~\cite{MN}.

As explained above the standard factorization procedure is
associated with poles at $u=0$ in the Borel representation of the
separate subprocesses. For the fragmentation function this is a
{\em logarithmic ultraviolet divergence} and for the coefficient
function a logarithmic divergence of collinear origin. With the
appropriate subtraction both functions become free of
$u\longrightarrow 0$ singularities. Thus, each of them has a
well-defined perturbative expansion (which is factorization scheme
dependent) to any order.  However, this expansion does not
converge owing to the factorial increase of the coefficients
induced by renormalons: In the Borel  representation infrared
renormalons show up as poles at positive $u$, which need to be
avoided when performing the integral. This leads to a {\em
power-suppressed} ambiguity.

\begin{figure}[htb]
\begin{center}
\epsfig{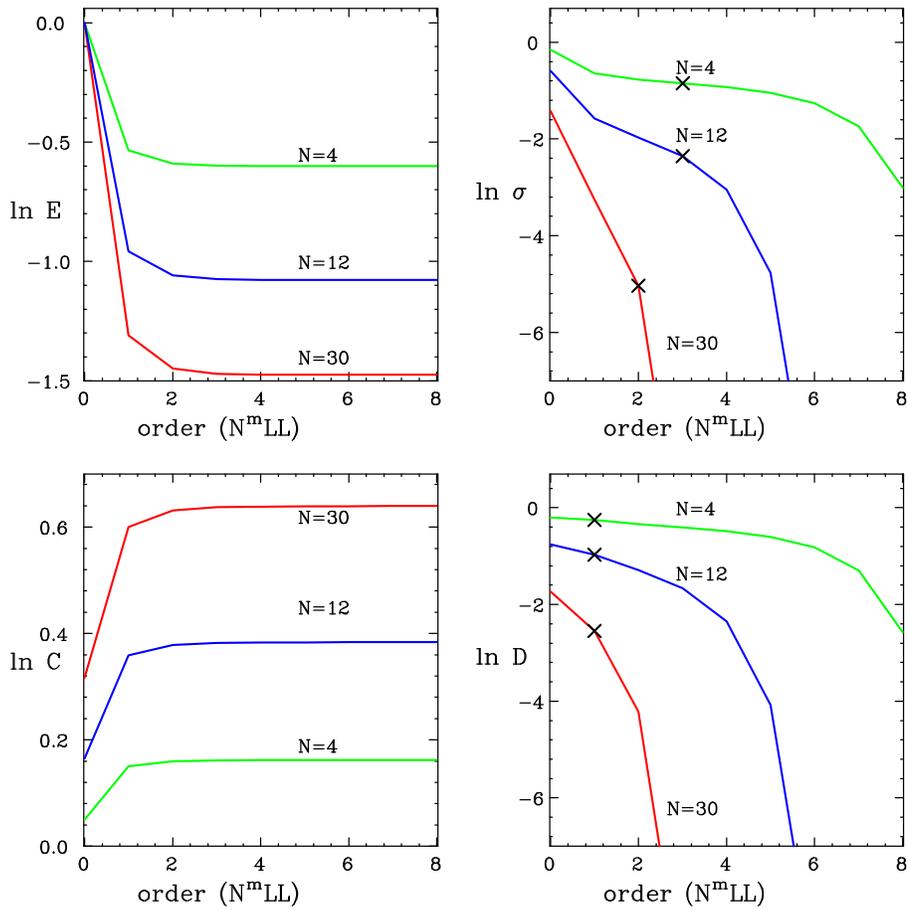}
\caption{\label{fig:logacc} 
Evaluation of the Sudakov exponents with increasing logarithmic accuracy.
The minimal term in the expansion is denoted by a cross.}
\end{center}
\end{figure}

Let us examine the nature of the perturbative expansion of the
Sudakov exponents. The DGE Sudakov exponent for the fragmentation
function $\ln \tilde{D}(N,m^2)$ of Eq.~(\ref{D_N_MSbar}), expanded
in powers of $A(m^2)$, is evaluated in Fig.~\ref{fig:logacc}
(lower right box) for $m = 4.75$~GeV, $\alpha_s^{\MSbar}(m^2) =
0.218$ and $N=4,\,12$ and $30$. The plot shows progressive partial
sums of a fixed-logarithmic accuracy, starting at LL, (m=0 in the
figure) and proceeding to higher logarithmic accuracy. It should
be noted that the result is exact only to NLL, while at higher
orders the values correspond to the DGE extrapolation from the
large-$\beta_0$ limit. The renormalon effect leading to divergence
of the series sets in early on: the minimal term is NLL (m=1) in
all three cases.

Similar plots of the Sudakov exponents  $\ln \tilde{J}(N,q^2)$, $\ln
\tilde{E}(N,q^2,m^2)$ and $\ln \tilde{\sigma}(N,q^2,m^2)$, of
Eqs.~(\ref{J_N}),~(\ref{E_N}) and~(\ref{sigma_N}), respectively, are shown in
the other boxes for a center-of-mass energy of  $\sqrt{q^2} = M_Z$ and
$\alpha_s^{\MSbar}(M_Z^2) = 0.118$.  Here the fixed-logarithmic accuracy
expansion is in powers of $A(M_Z^2)$. It is clear that the divergence of the
coefficient function is rather mild: $\ln \tilde{J}$ does not reach the minimal
terms up to m=8; $\ln \tilde{E}$ converges and the Sudakov exponent of the
total cross section $\ln \tilde{\sigma}$ reaches the minimal term at NNLL or at
N$^3$LL depending on $N$. Here the divergence is induced by the fragmentation 
subprocess.

When the expansion is truncated, e.g. at NLL order, the renormalon ambiguity
does not appear but power accuracy is usually not reached.
If one accepts that the DGE result~(\ref{D_N_MSbar}) represents well the
contribution of subleading logs, going to power accuracy
simply requires avoiding any arbitrary fixed-logarithmic accuracy
truncation.
Since the series does not converge, a sensible possibility is to define the
perturbative sum by truncation at the minimal term, i.e. just before the
series starts
diverging. The minimal term scales as a power.
This definition was recently used~\cite{Gardi:2002xm} in the context of
DGE for the deep inelastic structure function~$F_2$, where a purely
perturbative result was shown to be consistent with the data. However,
in the case of the
heavy-quark fragmentation considered here power corrections
are particularly large and the series starts diverging already at low orders.
In this case truncation at the minimal term would be far from optimal as it
leads to discontinuities in the perturbative result as a function of $N$ whose
magnitude is comparable to the power correction itself, thus inducing
${\cal O}(1)$
discontinuities in the non-perturbative contribution.
The divergence of the coefficient function $\tilde{C}(N,q^2)$ is much softer,
so here truncation at the minimal term would be numerically sensible.

A better regularization prescription is to evaluate the Borel
integral in Eq.~(\ref{D_N_MSbar}) directly, with a Principal Value
(PV) prescription applied at each pole. Fortran routines exist for
evaluating such integrals efficiently provided that the exact
position of the singularity is known, as in our case. This will
therefore be our default prescription. All the numerical results
for DGE we present below are obtained in this way.

\subsubsection{Comparison between DGE and NLL resummation}

We now compare the numerical results for NLL and DGE resummation
with PV regularization. Our default parameters will be $\sqrt{q^2}
= 91.2$~GeV, $m = 4.75$~GeV, $N_f = 5$ and $\Lambda = 0.243$~GeV,
corresponding to $\alpha_s^{\MSbar}(M_Z^2) = 0.118$ and
$\alpha_s^{\MSbar}(m^2) = 0.218$. We will be using a two-loop
running coupling\footnote{Note that the exact two-loop evolution
Eq.~(\protect\ref{eq:aslambert}) is close but not identical to the
commonly used expansion of the running coupling in $\ln \mu^2$,
(see e.g. Eq.~(29) in \cite{CC}). The choice we have made for
obtaining our default value $\alpha_s^{\MSbar}(M_Z^2) = 0.118$
corresponds to $\Lambda = 0.226$~GeV when Eq.~(29) in \cite{CC} is
used.} given by eqs.~(\ref{borelcoupling}) and (\ref{A_B}), or,
explicitly, by~\cite{Gardi:1998rf} \beq \alpha_s^{\MSbar}(\mu^2) =
\frac{\pi}{\beta_0}\left(-\frac{1}{\delta}\frac{1}{1+w(s)}\right)\;
, \label{eq:aslambert} \eeq where \beq
 w(s) \equiv
W_{-1}\left(-\exp(-s/\delta - 1)\right)\;,\qquad
s\equiv\ln\frac{\mu^2}{\Lambda^2}\;,\qquad
\delta\equiv\frac{\beta_1}{\beta_0^2}\;.
\eeq
Here $W$ stands for the Lambert $W$ function~\cite{lambert}.

\begin{figure}[thb]
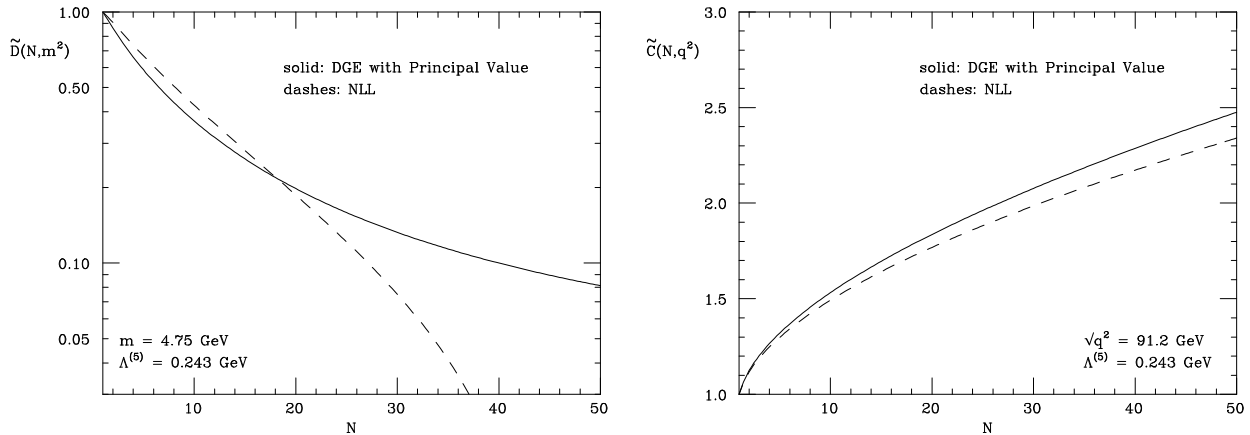

\begin{center}
\epsfig{file=dini-lambert.ps,width=8cm}~~~~
\epsfig{file=cf-lambert.ps,width=8cm}
\caption{\label{fig:nlldge} Left: the matched heavy-quark fragmentation function
$\tilde{D}(N,m^2)$  with
DGE (solid) and NLL (dashes) resummation.
Right: the same for the $e^+e^-$ coefficient
function $\tilde{C}(N,q^2)$.}
\end{center}
\end{figure}
Figure~\ref{fig:nlldge} shows a comparison between the NLL and DGE
resummed results for the heavy-quark fragmentation function
$\tilde{D}(N,m^2)$, and
for the~$e^+e^-$ coefficient function $\tilde{C}(N,q^2)$ (both are matched
 according to (\ref{eq:matching}) and (\ref{eq:matching-cf}) to the full ${\cal O} (\alpha_s$) result).
One can immediately see that the NLL result for $\tilde{D}(N,m^2)$
breaks down at $N = N_L =\exp(\pi/(2\beta_0\alpha_s(m^2))\sim 40$
 where the $g^{(i)}_{\rm ini}$ functions become singular.
 The DGE result with PV
regularization, on the other hand, remains moderate and shows no
singular behavior. Indeed, owing to the procedure adopted where
the Borel integrals are directly evaluated rather than expanded
the DGE result is
 free\footnote{Note that the convergence of (\ref{D_N_MSbar})
 at $u\longrightarrow \infty$ for any $N$ is guaranteed by the factor $\Gamma(-u)$.
 This factor also introduces an infinite set of renormalon singularities.}
of Landau singularities.

This does not mean, of course, that the {\em perturbative} DGE
result by itself is physically meaningful: power corrections are
needed. The advantage is, however, that this result, being free of
spurious singularities, provides a good basis for the
parametrization of power corrections for any $N$. In the case of
$\tilde{C}(N,q^2)$ the difference between the NLL and DGE results
is much smaller than for $\tilde{D}(N,m^2)$. The Landau
singularity in the Sudakov-resummed coefficient function is
located much further, at $N=\exp(\pi/(\beta_0\alpha_s(q^2))\sim
10^6$. Thus here subleading logs are smaller and power corrections
can definitely be neglected.

\begin{figure}[thb]
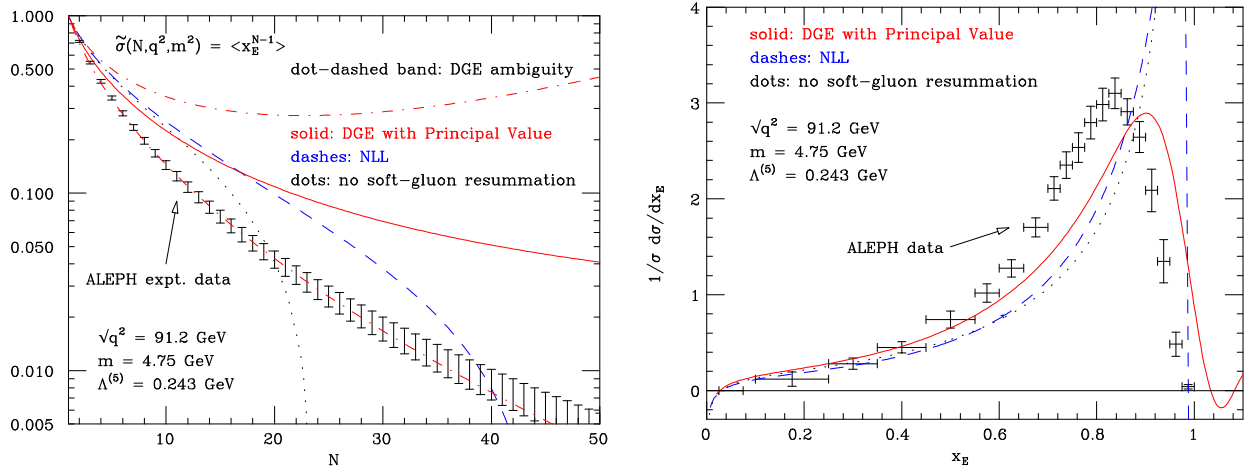

\begin{center}
\epsfig{file=full-lambert.ps,width=8cm}~~~~
\epsfig{file=full-x-lambert.ps,width=8cm}
\caption{\label{fig:nlldgefull} Left: a comparison between DGE and NLL for
heavy-quark production in the $e^+e^-$ process.
Right: the same curves in $x_E$ space.
ALEPH~\protect\cite{Heister:2001jg} data are also shown.}
\end{center}
\end{figure}

Figure~\ref{fig:nlldgefull} shows a similar comparison for the
full $e^+e^-$ cross section $\tilde{\sigma}(N,q^2,m^2)$ of
Eq.~(\ref{eq:eefull}). For reference we show also a curve with no
soft gluon resummation as well as the experimental data from
ALEPH~\cite{Heister:2001jg}.
The theoretical curves in $x_E$ space are obtained via numerical Mellin inversion:
\be
D(x) = \frac{1}{2\pi i} \int_{\cal C} dN\; x^{-N} \tilde{D}(N) \; .
\ee
\begin{figure}[htb]
\begin{center}
\epsfig{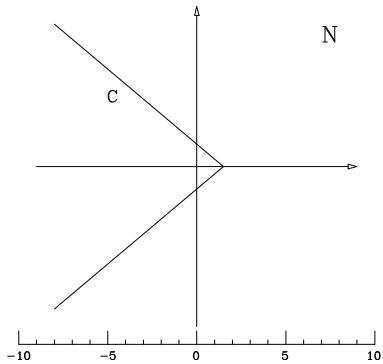}
\caption{\label{fig:contour} A prototype of the
 contour ${\cal C}$ employed in the
numerical Mellin inversion to $x$ space.}
\end{center}
\end{figure}
In the NLL case the Minimal Prescription~\cite{Catani:1996yz} is
used: the contour ${\cal C}$ (see Fig.~\ref{fig:contour})
passes between the origin and the leftmost Landau pole.
In the case of the (PV-regularized) DGE result a specific prescription 
is not required since there are no Landau singularities.
Any contour passing to the right of the origin would yield the same
result.
This may be surprising at first sight, since the original expression
for the fragmentation function in $x$ space~(\ref{ren_sum}) does have a
convergence constraint (see for
example~(\ref{1loop_converg_constraint})). 
However, as we saw, when going to moment space this
constraint is traded for renormalon singularities. Landau 
singularities instead appear only if the resummed expression is {\em expanded} 
in terms of the coupling. 
The inherent ambiguity of the perturbative description of the fragmentation 
function as \hbox{$x\longrightarrow 1$} appears in the Borel sum as a 
renormalon ambiguity:
In order to compensate for it one needs to specify an infinite set of power
terms~$(N \Lambda /m)^n$. 

A hint about the size of such contributions can be obtained by
considering different prescriptions for performing the
Borel integration, PV being one of the possible choices. Evaluating $\ln
\tilde{D}(N,m^2)$ in (\ref{D_N_MSbar}) by going above and below
the {\em first} pole at $u=1/2$ we obtain the band bounded by the
dot-dashed curves in Fig.~\ref{fig:nlldgefull}. The experimental data
turn out to be within the band: This supports our expectation that
one can describe them by modifying the
resummed perturbative result with the first few power corrections.

\subsection{Non-perturbative contributions}
\label{sec:npfit}

In general, the non-perturbative contribution to the fragmentation
function should be included on top of the perturbative one through
a convolution in $x$. Moments of the $B$-meson cross section as
measured at LEP can then be written as the product of the
perturbatively calculated moments
$\tilde{\sigma}^{\PT}(N,q^2,m^2)$ and the moments of some
non-perturbative function.

Our first result is that at large $N$ and $m$ this function
depends on the ratio $N\Lambda/m$ alone, up to corrections of
order $1/N$. Unfortunately, with just one heavy flavour (charm is
probably too light) the predictive power of this statement is
largely lost: fixing $m$ the fragmentation can anyway depend only
on $N$. This situation can be contrasted, for example, with the
case of event-shape distributions where non-perturbative
corrections can a priori depend on the centre-of-mass energy as
well as on the shape variable: Thus the statement that the
corrections can be described by a shape function of a single
argument is already quite constraining~\cite{KS,Gardi:2002yg}.

The emphasis in the phenomenological analysis is therefore on the
particular dependence of the fragmentation function on
$N\Lambda/m$ which is predicted by the renormalon
model~(\ref{D_NP}). Let us write 
\beq \tilde{\sigma}(N,q^2,M^2)
\simeq \tilde{\sigma}^{\PT}(N,q^2,m^2)\,
\tilde{D}^{\NP}_{\{\epsilon_n\}}((N-1)\Lambda/m) 
\label{eq:pt-np}
\eeq 
where 
\beq \tilde{D}^{\NP}_{\{\epsilon_n\}}((N-1)\Lambda/m) =
\exp\left\{- \sum_{n=1}^\infty \epsilon_n
\left(\frac{(N-1)\Lambda}{m}\right)^n \right\}. 
\label{eq:np} 
\eeq
The separation being, of course, ambiguous, the perturbative part
in Eq.~(\ref{eq:pt-np}) describes the production of a bottom quark
surrounded by a cloud of soft gluons and the non-perturbative one
describes its hadronization into an observable $B$ meson. As before, we
have not explicitly shown the dependence on the QCD scale $\Lambda$
which the perturbative part  $\tilde{\sigma}^{\PT}$ acquires via the
strong coupling. The
dependence of $\tilde{\sigma}(N,q^2,M^2)$ on the heavy meson mass
$M$ rather than the heavy quark mass $m$ serves as a reminder that
it refers to the observed particle.

In Eq.~(\ref{eq:np}) we have made a few modifications compared to
Eq.~(\ref{D_NP}). Firstly, the factor $C_F/\beta_0$ has been
included into the parameters $\epsilon_n$. Secondly, we have
introduced a term proportional to $(N\Lambda/m)^2$, which is
absent in the renormalon result. By including it and fitting to
data we can examine this feature of the renormalon
model~(\ref{D_NP}). Finally, we have replaced $N\to N-1$. This
modification is of course allowed in the large-$N$ limit, and it
implements the constraint that for $N=1$ (i.e. the normalized
total cross section) the non-perturbative correction vanishes.
Eq.~(\ref{eq:pt-np}) will be the master equation for our
phenomenological analyses of the LEP data. The default choice for
the perturbative calculation of the $b$-quark differential cross
section $\tilde{\sigma}^{\PT}(N,q^2,m^2)$ will be the
PV-regularized DGE.

\subsubsection{The experimental data}

Several experimental collaborations have recently published
high-statistics and high-accuracy data for $B$-meson production in
$e^+e^-$ collisions at the $Z^0$ peak, i.e. centre-of-mass energy
around 91 GeV. Most of the data have been published in the form of
differential distributions in the variable $x_E$, representing the
ratio between the $B$-meson energy in the laboratory frame and the
beam energy. The formal definition of the fragmentation
function~(\ref{D_def}) is in terms of the longitudinal momentum
fraction $x$. However, for large $q^2$ the two coincide:
$x_E\simeq x$ up to corrections (\ref{Ep1}) of order $m^2/q^2$
which we neglect. From here on the two notations will be used
interchangeably.

Data in $x$ space, along with their covariance matrices, have been
published by the SLD~\cite{Abe:2002iq},
ALEPH~\cite{Heister:2001jg}, DELPHI~\cite{delphi} and
OPAL~\cite{Abbiendi:2002vt} Collaborations. The DELPHI
Collaboration has also published moment data, including the very
important correlations between the moments up to $N=6$.

Rather than converting our moment-space expressions to $x$ space
as usually done, we shall perform our fits directly in moment
space, using the CERN Library minimization routine 
\verb+MINUIT+~\cite{James:1975dr}.
Given the $x$-space data and covariance matrices as inputs,
moments and their own covariance matrices can be calculated. The
integral~(\ref{D_mom}) defining Mellin moments can be replaced by 
a discrete sum
over all the bins in $x$ space ($n_{\bins}$ in total), 
\beq
\tilde{\sigma}(N,q^2,M^2)  \simeq \sum_{i=1}^{n_{\bins}} x_i^{N-1}
\Delta x_i f_i, \label{discretemom} 
\eeq 
where $f_i$ is the normalized value of the $i$-th bin, $x_i$ its
central abscissa and $\Delta x_i$ its width. This equation
represents a linear transformation of the vector $f_i$, 
\beq
\underline{\tilde{\sigma}} = {\bf A}\;\underline{f}\; , 
\eeq 
and the matrix $\bf A$ is defined by $A_{Ni}\equiv x_i^{N-1} \Delta
x_i$. Given the covariance matrix $\bf F$ for the $f_i$ values,
one can build the covariance matrix $\bf S$ for the
$\tilde{\sigma}(N,q^2,M^2)$ moments as \beq {\bf S} = {\bf
A}\;{\bf F}\;{\bf A}^T \; . \label{eq:covmat} \eeq

\begin{table}[thb]
\begin{center}
\small
\begin{tabular}{|c|c|c|c|}
\hline
$N$ & ALEPH & DELPHI~\cite{delphi,eli} & DELPHI \\
\hline
\hline
2 &0.7163 $\pm$ 0.0085 &0.71422 $\pm$ 0.0052  & 0.7147  $\pm$ 0.0045\\
3 &0.5433 $\pm$ 0.0097 &0.5401 $\pm$ 0.0064  & 0.5413  $\pm$ 0.0057 \\
4 &0.4269 $\pm$ 0.0098 &0.4236 $\pm$ 0.0065  & 0.4248  $\pm$ 0.0060 \\
5 &0.3437 $\pm$ 0.0096 &0.3406 $\pm$ 0.0064  &0.3419  $\pm$ 0.0059 \\
6 &0.2819 $\pm$ 0.0094 &0.2789 $\pm$ 0.0061  &  0.2804  $\pm$ 0.0057\\
\hline
7 &0.2345 $\pm$ 0.0091 & -                   & 0.2333  $\pm$ 0.0054\\
8 &0.1975 $\pm$ 0.0087 & -                   & 0.1965  $\pm$ 0.0050\\
9 &0.1680 $\pm$ 0.0084 & -                   & 0.1672  $\pm$ 0.0047\\
10 &0.1441 $\pm$ 0.0080 & -                  &0.1435  $\pm$ 0.0044\\
11 &0.1245 $\pm$ 0.0076 & -                   & 0.1241  $\pm$ 0.0041\\
12 & 0.1084 $\pm$ 0.0072&   -                & 0.1081  $\pm$ 0.0038\\
13 & 0.0949 $\pm$ 0.0069&   -                &  0.0947  $\pm$ 0.0036\\
14 & 0.0835 $\pm$ 0.0065 &   -               & 0.0835  $\pm$ 0.0033  \\
15 & 0.0738 $\pm$ 0.0062  &   -              & 0.0740  $\pm$ 0.0031  \\
16 & 0.0656 $\pm$ 0.0058  &   -              & 0.0659  $\pm$ 0.0029  \\
17 & 0.0585 $\pm$ 0.0055  &   -              &  0.0589  $\pm$ 0.0027 \\
18 &  0.0524 $\pm$ 0.0052 &   -              &  0.0529  $\pm$ 0.0026 \\
19 & 0.0471 $\pm$ 0.0049  &   -             & 0.0477  $\pm$ 0.0024  \\
20 & 0.0425 $\pm$ 0.0047  &   -             & 0.0432  $\pm$ 0.0023  \\
21 &  0.0384 $\pm$ 0.0044 &   -             &  0.0393  $\pm$ 0.0022 \\
\hline
\end{tabular}
\end{center}
\caption{\label{table1}Experimental data for normalized moments of
weakly-decaying $B$-meson
production in $e^+e^-$ collisions from the ALEPH and
DELPHI Collaborations at LEP, evaluated according to
Eq.~(\ref{discretemom}) (left and right columns). 
The set in the central column
corresponds to the five moments published by DELPHI in
\protect\cite{delphi} (the $N=2$ point does not include here the
correction for initial state electromagnetic radiation~\protect\cite{eli}).}
\end{table}

Table~\ref{table1} contains most of the moment-space data which will be
used in the fits.
Moments constructed from ALEPH and DELPHI data are
shown.
Moments higher than those shown in Table~\ref{table1} can of course
be calculated, and they have also been
used in some of the fits. The amount of independent information they carry
is however limited.

The central values reported in Table~\ref{table1} show a very good agreement
between the ALEPH and DELPHI sets. 
 DELPHI moments are reported twice in the Table. The data
in the central column are directly provided by \cite{delphi,eli}, where
they were published in preliminary form, 
whereas we have calculated the ones in the right one according to
Eq.~(\ref{discretemom}).  It should be noted that the DELPHI
Collaboration extracted the moments directly from the unfolding
program they used, rather than calculating them by summing up the
published bins in $x$ space. This accounts for the small
differences. The two sets of moments are, however, compatible
within errors.  For full consistency, we shall make use of the
published set of moments with the published covariance matrix, and
of our own values together with a covariance matrix we extract
ourselves according to Eq.~(\ref{eq:covmat}). The latter also
presents some differences with the published one.

As a test of the fitting program, we have reproduced the fit that
the DELPHI Collaboration performed in \cite{delphi} to the five
correlated moments which were published there, employing a
so-called Kartvelishvili \cite{kart} functional form,
$D^{\NP}(\alpha,x) = (\alpha+1)(\alpha+2) x^\alpha (1-x)$. Using
their parameters, their moments and covariance matrix, and NLL
resummation only, we do recover their result: $\alpha = 17.07 \pm
0.57$, $\chi^2 = 114.7$ for $5-1$ degrees of freedom. However,
fitting the moments we calculate ourselves (right column in
Table~\ref{table1}) and using our own covariance matrix, we get
instead $\alpha = 14.30 \pm 0.37$ and $\chi^2 = 159.8$. The
apparent incompatibility of these results should mainly be
attributed to the intrinsic failure of the fit, as indicated by
the very large $\chi^2$ values. It also shows, however, how strong
is the effect of small variations in the moments and in their
covariances, due to the high degree of correlation and to the very
small experimental errors.
We have also performed the same fit to the first five ALEPH moments, obtaining
$\alpha = 21.5 \pm 2.0$ and $\chi^2 = 15.1$. 

We wish to point out that the very strong correlations
between the experimental data for the moments are an important
feature, which cannot be neglected when fitting a hadronization
model. Extracting a meaningful covariance matrix for many moments
at once is however no straightforward task. Because of the strong
correlation between successive moments, inversion of such a matrix
-- needed when evaluating the $\chi^2$ for the fits -- quickly
becomes an intractable problem. In practice the numerical accuracy
issue becomes acute when more than five or six consecutive moments
are fitted together. Possible solutions are to use only a few
moments or to pick non-consecutive moments. By means of these
choices the matrix becomes smaller and the degree of correlation
is lessened. One observes that matrices containing five or six
moments, taken two or three moments apart, can be inverted with
acceptable precision up to $N\sim20$ . Using only three or four
moments  a sufficient numerical accuracy can be obtained up to
$N\sim 30$. We make use of this expedient in some of the fits
described below.

\subsubsection{Fits to low moments}

In order to test the renormalon model for the non-perturbative
fragmentation function we would like to fit at least three
parameters:
 $\Lambda$, $\epsilon_1$ and~$\epsilon_2$. $\epsilon_1$ is
the leading power correction  corresponding to the shift of the
entire distribution in $x$ space in units of $\Lambda/m$. It is
expected to be of order $1$. Based on the renormalon
model~(\ref{D_NP}), $\epsilon_2$ is expected to vanish. Note that
varying $\Lambda$, i.e. the value of $\alpha_s^{\MSbar}(M_Z^2)$, 
influences both the perturbative and the 
non-perturbative parts of the cross section (\ref{eq:pt-np}). 
Our approach to heavy-quark fragmentation can thus be tested by checking that the fit returns a value which is consistent with other determinations of
$\alpha_s^{\MSbar}(M_Z^2)$.

With this task in mind and the data described above at hand we should
choose a subset of moments to be fitted. Theoretical limitations exist
both for very small $N$ and for very large $N$. In the former case
neither the  soft-gluon perturbative resummation nor the power
corrections of~(\ref{D_NP}) are expected to dominate. In the latter,
parametrizing just a few power corrections in the exponent 
may not be sufficient, as $N\Lambda/m$ is not small.

In this section we concentrate on moments up to $N=6$. 
In the next section we extend the analysis to higher moments. We use the
correlated moments obtained from the ALEPH
data and shown in Table~\ref{table1}.  DELPHI moments have also been used for
cross checks and comparisons but, due to their preliminary nature, we 
refrain from quoting these results. 

\begin{table}[ht]
\begin{center}
\begin{tabular}{|l|c|c|c|}
\hline
&\multicolumn{3}{c|}{$\epsilon_n = 0 \,\, {\rm for}\,\, n \geq 3$}\\
\hline
$\epsilon_1$    & 1.23 $\pm$ 0.23    & 1.03 $\pm$ 0.13   & 2.9 $\pm$ 1.5\\
$\epsilon_2$    & 0 (fixed)          & 0.011 $\pm$ 0.095 & -1.1 $\pm$ 1.2\\
$\Lambda$ (GeV) & 0.217 $\pm$ 0.025  & 0.243 (fixed)     & 0.138 $\pm$ 0.045\\
$\chi^2$/d.o.f  &  6.8/3             &  7.9/3            & 3.7/2 \\
\hline
\end{tabular}
\caption{\label{table2}Results of fits 
to ALEPH moments from $N=2$ to $N=6$. }
\end{center}
\end{table}

\begin{figure}[htb]
\begin{center}
\epsfig{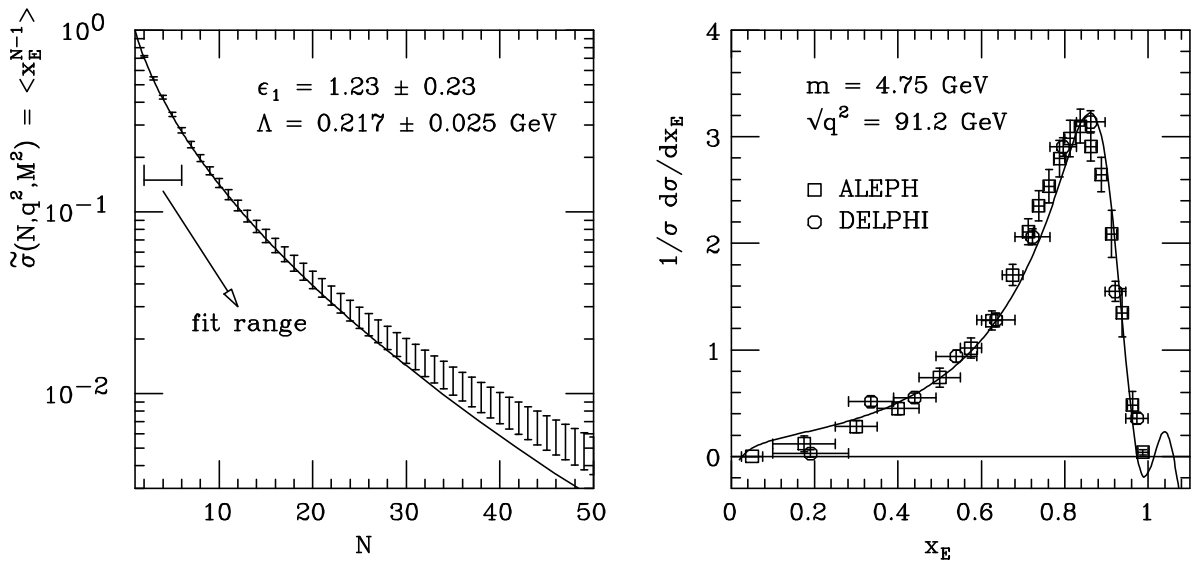}
\end{center}
\caption{\label{fig:shiftcurves} Left: results of a fit
for $\Lambda$ and $\epsilon_1$ (setting $\epsilon_n = 0$ for $n\geq 2$) 
to ALEPH moments $N=2$ to $6$.
Right: the corresponding curve in $x_E$ space, compared to the ALEPH
and DELPHI data.}
\end{figure}

First, we wish to extract $\Lambda$ and
test the prediction that the leading power correction is of the
order of $\Lambda/m$. We do this by fitting $\Lambda$ and
$\epsilon_1$, setting $\epsilon_n = 0$ for any $n\geq 2$ in
Eq.~(\ref{eq:np}). The results of this fit are shown in the first column
of Table~\ref{table2}.
The
$\chi^2$ of the fit is reasonable albeit somewhat large.
It is important to
notice that the central value for $\Lambda$, corresponding to
$\alpha_s^{\MSbar}(M_Z^2) \simeq 0.117$, is fully compatible
with present determinations of the strong coupling. The best fit
value for $\epsilon_1$ is of order~$1$, as expected.
The resulting curves and the ALEPH data are shown in 
Fig.~\ref{fig:shiftcurves}, both in $N$ and
$x_E$ space.
The overall description of the data is reasonable. 
This is quite remarkable given that, in $x_E$ space, the theoretical curve 
is no more than a shift of the PV-regularized DGE perturbative result. 

The self-consistency of this fit can be probed by fitting now
$\epsilon_2$ while fixing $\Lambda = 0.243$~GeV. Being this value 
for $\Lambda$ compatible with the one returned by the previous fit,
we expect to get a compatible value for $\epsilon_1$. The results are
shown in the second column of Table~\ref{table2}.
$\epsilon_1$ is indeed consistent with the one
shown in the first column, the $\chi^2$ is similar, and $\epsilon_2$
turns out to be consistent with zero.

Finally, we attempt fitting three parameters,  $\epsilon_1$,
$\epsilon_2$ and $\Lambda$, simultaneously.  
The results are shown in the third column of Table~\ref{table2}.
Consistency with previous fits appears rough, and the errors are
extremely large.

Because of the problems encountered in the three-parameter fit,
and since we expect the perturbative corrections we resum and the
power correction we parametrize to become dominant at larger $N$,
we now proceed to analyze higher moments.

\subsubsection{Fits to high moments}

In this section we wish to explore the possibility of fitting
Eq.~(\ref{eq:pt-np}) to
large-$N$ correlated moments. The large-$N$ region is challenging
in several respects. Considering the data, the finite binning in
$x_E$-space limits the amount of independent information contained
in large-$N$ moments. On the theoretical side, as soon as the
condition $N\Lambda/m\ll 1$ is violated the non-perturbative
contribution is expected to become comparable to, and eventually to
dominate, the perturbative one. Moreover, the {\em
exponentiation} of power corrections can no longer be ignored and
the specific form of~(\ref{D_NP}), e.g. the {\em vanishing of the
second power} of $N\Lambda/m$, becomes relevant. This makes this
exercise particularly interesting: The validity of our model can
really be tested. Yet, it is a priori not known up to what value
of $N$ Eq.~(\ref{D_NP}) with just a few parameters might hold: One
should keep in mind that when going to extremely large $N$ the number
of relevant parameters, corresponding to increasing powers of
$N\Lambda/m$ in the exponent, increases, and the fitting procedure
may get out of control.

\begin{table}[htb]
\begin{center}
\begin{tabular}{|l|c|c|c|c|}
\hline
$N$& $6,8,10,12,14,16$ & $13,15,17,19$ &  $7,10,13,16,19$ &
$21,24,27,30$ \\
\hline
\hline
$\epsilon_1$                  &   0.831 $\pm$ 0.041  & 0.77 $\pm$ 0.15
& 0.96 $\pm$ 0.12 & 0.817 $\pm$ 0.056\\
$\epsilon_2$                  &   -0.037 $\pm$ 0.007 & -0.024 $\pm$0.028
& -0.063 $\pm$ 0.025 & -0.040 $\pm$ 0.008\\
$\Lambda$ &   0.261 $\pm$ 0.015  &  0.255 $\pm$0.025
& 0.238 $\pm$ 0.022 & 0.256 $\pm$ 0.020\\
$\chi^2$/d.o.f                &   11/3             & 8.3/1
&9.0/2                       & 3.3/1\\
\hline
\end{tabular}
\caption{\label{table9}Fits to correlated ALEPH moments in the large-$N$
region. Only a few non-consecutive moments are used, as indicated in the
various columns.}
\end{center}
\end{table}

The moments we fit here are those we reconstructed from the
$x_E$-space data and covariance matrices\footnote{We use covariance
matrices with more significant figures~\cite{boccali} than published
in~\cite{Heister:2001jg}.}
published by ALEPH. 
The fine binning in $x_E$ space of the ALEPH data helps
in providing a good description of the large-$N$ moments in which
we are interested\footnote{For a given binning, at extremely large $N$ the
moments are eventually determined only by the very last bin at large
$x$, and thus higher moments carry no new information even in an ideal
situation where the bins are not correlated. This is not yet
realized for the ALEPH data at $N$ of a few tens. For example, for
$N=10$ there are twelve bins that contribute more than the last
one, for $N=20$ this number drops to seven and for $N=30$ there
are still five such bins.}. Provided that the correlations are properly
taken into account, it allows one to consider moments as high as
$N\simeq 30$.

In order to achieve a sufficiently\footnote{Note that the fits 
in the large-$N$ region are performed at
the edge of the numerical accuracy permitted by the inversion of the
covariance matrix.}
accurate inversion of the covariance
matrix we now make use of the expedient of using non-consecutive
moments. The results, shown in Table~\ref{table9}, are remarkably
consistent: Over a wide range of moments the best-fit values for
$\Lambda$, $\epsilon_1$ and $\epsilon_2$ appear very similar, and
in line with our expectations.
\begin{figure}[htb]
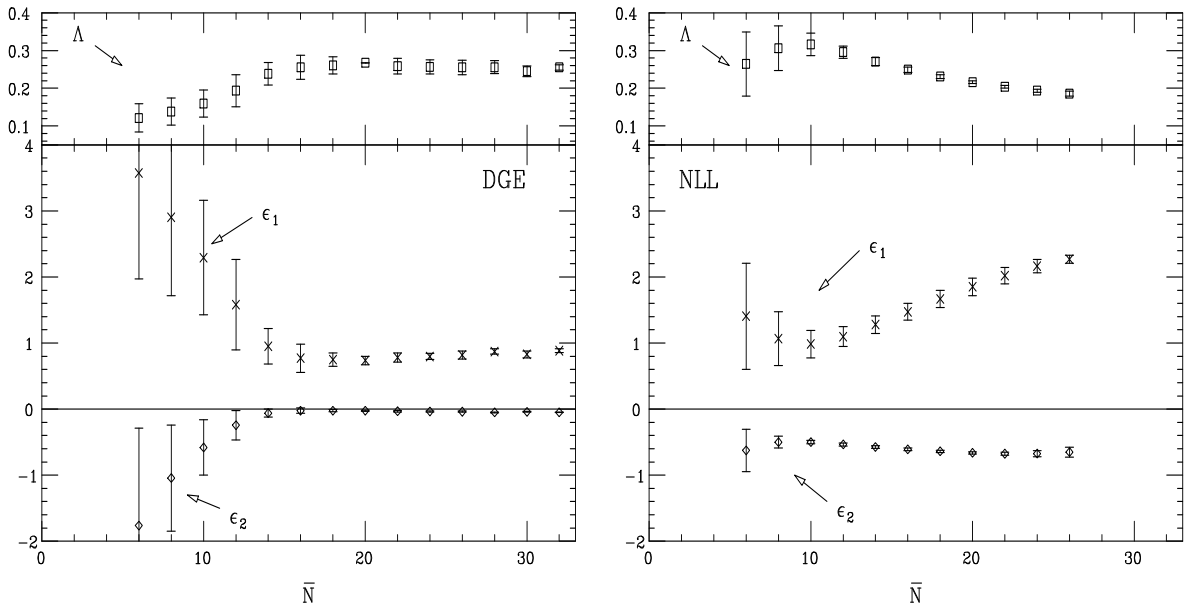

\begin{center}
\epsfig{file= roll-aleph-3par-step2-bin3.ps,width=7.6cm,height=8cm}
\hspace{.2truecm}
\epsfig{file= roll-aleph-3par-step2-bin3-nll.ps,width=7.6cm,height=8cm}
\end{center}
\caption{\label{fig:roll-correlated} Results of fits to  $\epsilon_1$, 
$\epsilon_2$ and $\Lambda$ (in GeV on the vertical
axis) performed with correlated ALEPH moments within a ``rolling
window'', centered at $\bar{N}$, where 4 moments out of 7 are
used. In the left plot the DGE perturbative result (with PV
regularization) is used. In the right one instead the resummation
is truncated at NLL accuracy.}
\end{figure}

We further explore the large-$N$ region by employing `rolling window'
fits, using again non-consecutive moments.
Figure~\ref{fig:roll-correlated} shows fits to $\Lambda$, $\epsilon_1$
and $\epsilon_2$, performed within windows bordered by $\bar{N} \pm 3$,
where only four correlated moments are used. At small $N$ the fits are
inconclusive. As we saw above (Table~\ref{table2}) this is a
consequence of trying to fit too  many parameters. In the large-$N$
region the situation improves dramatically:  $\epsilon_2$ tends to
zero, as predicted by the renormalon model, and  the value for
$\Lambda$ stabilizes on \hbox{$\sim$ 0.25 $\pm$ 0.04 GeV},
corresponding to $\alpha_s^{\MSbar}(M_Z^2) \simeq 0.118 \pm 0.003$,
which is compatible with other determinations. We stress that this
should not be regarded as a reliable determination of the strong
coupling as the analysis of experimental and theoretical errors was not
performed. Note the fluctuations in the error bars at large $\bar N$
values, due to the deteriorating numerical accuracy.

For comparison, Fig.~\ref{fig:roll-correlated} also shows the same
kind of fits with NLL resummation. The range in $N$ is limited
here by the non-physical behavior of the resummed result for $N$
greater than 30 or so, as shown in Fig.~\ref{fig:nlldgefull}. The
results are clearly at variance with the DGE fits: $\Lambda$ and
$\epsilon_1$ do not stabilize and $\epsilon_2$ does not tend to
zero when $N$ becomes large.

\subsubsection{Comparison with moment-space and $x_E$-space data}

At the end of the day, one wishes of course to extract a
parametrization for non-perturbative effects such that, when
convoluted with the appropriate perturbative contribution, it
allows for a fair description of the experimental data in both
moment and $x_E$ space.

\begin{figure}[htb]
\begin{center}
\epsfig{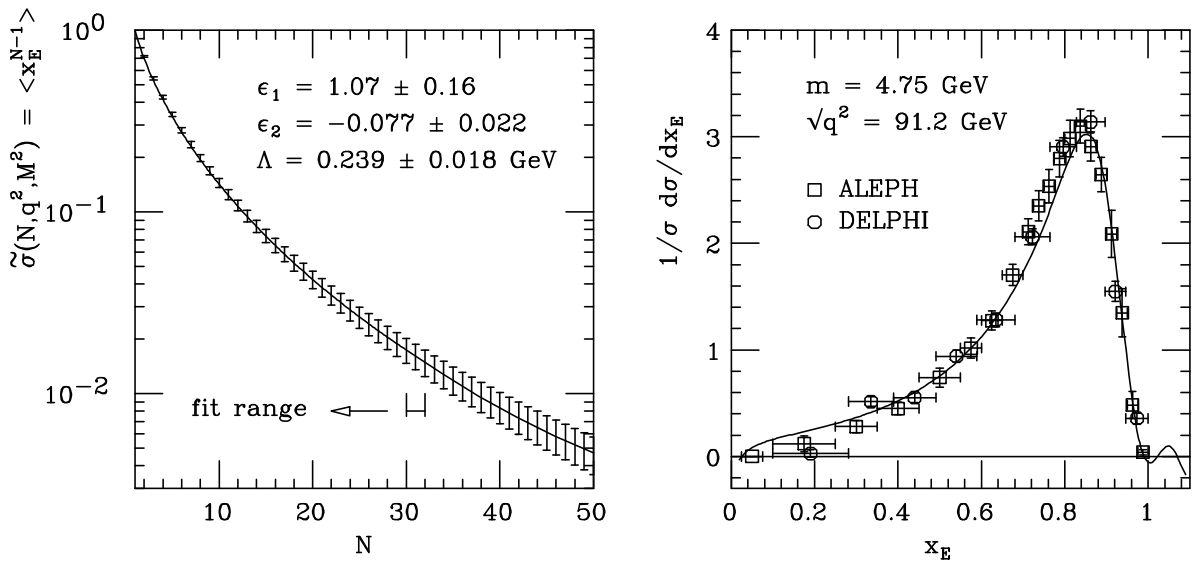}
\end{center}
\caption{\label{fig:curves} Left: results of a fit
for $\Lambda$, $\epsilon_1$ and $\epsilon_2$ and
 to ALEPH moments $N=30,31,32$, setting $\epsilon_n = 0$ for $n\geq 3$.
Right: the corresponding curve in $x_E$ space, compared to the ALEPH
and DELPHI data.}
\end{figure}
The experimental
papers~\cite{Abe:2002iq,Heister:2001jg,delphi,Abbiendi:2002vt} have
tested a number of different parametrizations, and established which
ones seem to offer a good description of the data. Our present attempt
differs from theirs in several respects:
\begin{itemize}
\item we only fit data in {\sl moment space}, and subsequently derive
the corresponding $x_E$ distribution;
\item our perturbative description is based on DGE,
matched to the NLO result (\ref{eq:matching}), and to the NLL
Altarelli-Parisi evolution~(\ref{eq:eefull}). The experiments
instead usually use the perturbative fragmentation provided by
Monte-Carlo programs. We recall that a non-perturbative function
should only be used in connection with the same perturbative
description it has been fitted with. Conversely, changing the
perturbative description (or its parameters) can lead to fit vastly
different non-perturbative functions;
\item we use a renormalon-motivated functional form, Eq.~(\ref{eq:np}),
for the non-perturbative function, rather than phenomenological
models like the commonly used Peterson et al.~\cite{peterson} or
Kartvelishvili et al.~\cite{kart}. Besides the theoretical reasons
for using Eq.~(\ref{eq:np}), we have also checked that the
Kartvelishvili or Peterson functional forms do not provide a good
description of the data when used with our DGE-improved
perturbative result.
\end{itemize}

For illustrative purposes we perform a further fit using
exclusively very high moments, $N=30,31$ and $32$. 
The main disadvantage of such a fit is that higher power corrections on 
the exponent, which are not parametrized but simply set to zero, may in fact 
play some role. On the other hand it has several advantages:
\begin{itemize}
\item one can be confident that the perturbative
corrections resummed by~(\ref{D_N_MSbar}) and the power correction
parametrized by~(\ref{D_NP}) indeed dominate.
\item the stability observed in our previous fits at large $N$ guarantees
 that the result will be independent of the particular set of moment chosen.
\item it ensures a good description of the large-$N$ limit, and thus of the
region near $x\longrightarrow 1$.
\end{itemize}
 
The results are shown in Fig.~\ref{fig:curves}. The fitted
parameters (see the figure) are compatible with the ones shown in
Table~\ref{table9} and in Fig.~\ref{fig:roll-correlated}. $\epsilon_2$,
although not vanishing within the errors, is very small.
In order to check the sensitivity to higher power corrections in the
exponent, we have performed a similar fit (using $N=30,31$ and $32$)
fixing $\epsilon_2\equiv 0$ and having $\Lambda$,  $\epsilon_1$ and
$\epsilon_3$ as free parameters. The result is visually similar to
Fig.~\ref{fig:curves}, but the parameters are somewhat different:
$\Lambda=0.266 \pm 0.020$, corresponding to 
$\alpha_s^{\MSbar}(M_Z^2)\simeq 0.120$, and the parameters of the  shape
function are $\epsilon_1 = 0.86 \pm 0.11$ and  
$\epsilon_3=-0.011 \pm 0.003$. These differences with respect to
Fig.~\ref{fig:curves} represent the sensitivity to  higher power
corrections.

Despite
having used large-$N$ data only in the fit, the low-$N$ moments are fairly
well described by the resulting curve, typically deviating by no
more than a few percent\footnote{It is worth noting that the level
of accuracy of a few percent on the low-$N$ moments is by far
sufficient for phenomenologically relevant applications such as
the description of heavy-quark hadroproduction as performed
in~\cite{Cacciari:2002pa}.}. One can moreover notice that the
whole curve in $x_E$ space is well reproduced, notwithstanding the
fact that it was never directly fitted: Its shape is produced
by the DGE-improved perturbative calculation, properly
shifted and modeled by the fitted non-perturbative term (see also
Fig.~\ref{fig:nlldgefull} for a comparison with the purely
perturbative result). A better description of the lower moments
and the region left of the peak is expected upon including
corrections of order ${\cal O}((m^2/q^2)\alpha_s)$,
non-logarithmic ${\cal
O}(\alpha_s^2)$ corrections~\cite{Nason:1999zj}, and power corrections of
order $\Lambda/m$ not enhanced by~$N$.

Note that the curve in $x_E$ space remains positive up to $x_E$ very close to 
one, and beyond the last experimental point. 
This constitutes a marked improvement with respect to fits based on
fixed-order or NLL Sudakov-resummed perturbative results. In these
approaches a convoluted non-perturbative function can be fitted to the 
data, and the
resulting curve describes well the experimental points 
in the region below the peak.
However, the  cross section turns out negative over a 
broader region between the peak and $x_E=1$, and cannot therefore
describe the last few data points.
We recall that this improvement has been obtained here 
by {\sl refining} the perturbative prediction, adding
leading and subleading Sudakov logarithms within the DGE formalism and
introducing an appropriate prescription\footnote{Although the
prescription itself is arbitrary, the ambiguity is of a specific
functional form, namely that of the power corrections we introduce.
Thus, upon performing a fit where these corrections are parametrized
the ambiguity is fully canceled~\cite{Gardi:1999dq}: other
regularizations simply correspond to a redefinition of the parameters
which control power terms.} (PV) to regularize the renormalons.  This
stands in contrast with the approach of~\cite{bkk} where large
logarithmic corrections in the function $D(x,m^2)$ are simply
discarded.

\section{Conclusions}

In this paper we considered the problem of heavy-quark
fragmentation in the large-$x$ region. We first derived a general
relation (\ref{large_N}) between the fragmentation function at
large~$N$ and a non-local hadronic matrix element at large
light-cone separations. When combined with the result on the
asymptotic form of this matrix element at large $m$~\cite{JR}, it
implies that the {\em simultaneous limit} where $N$ and $m$ get
large together, with the ratio between them fixed, has a special
status: the asymptotic heavy-quark fragmentation function depends
on this ratio alone.

Equipped with this exact asymptotic result we proceeded to
evaluate the fragmentation function perturbatively, resumming
large logarithms of $N$ as well as running-coupling effect. We
first calculated the splitting function of an off-shell gluon of a
massive quark, from which we derived an all-order result for the
fragmentation function in the large-$\beta_0$ limit. This result
has some interesting features: in~$x$ space it does not contain
any renormalons, but it does have convergence constraints for
$x\to 1$. In moment space these constraints are
replaced by an infinite set of infrared renormalons, indicating
certain non-perturbative corrections to the fragmentation function.
The physical interpretation of this result is clear: the perturbative 
calculation only gives the probability of producing an on-shell
quark with an energy fraction $x$ 
surrounded by a cloud of soft gluons, whereas the physical
quantity (defined in the full theory) is the probability 
to produce a bound state (a heavy meson) with that energy 
fraction. The gap between the two is filled by these
 power corrections.

As expected, at large $N$ renormalon ambiguities depend only on the ratio~$N
\Lambda/m$. Thus, the non-perturbative contribution to the
fragmentation function at large $x$ can be incorporated through a
convolution with shape function of $m(1-x)$. Further details on
this function can be deduced from the structure of the perturbative result:
\begin{itemize}
\item similarly to the logarithmically enhanced contributions, these power 
corrections {\em  exponentiate} in moments space;
\item the leading power correction is ${\cal O}(N\Lambda/m)$.
Upon exponentiation it generates a shift of the perturbative
spectrum in $x$ by an amount proportional to $\Lambda/m$;
\item  the second corrections in the exponent, ${\cal O}(N^2\Lambda^2/m^2)$, is absent,
suggesting that the shift may be a good approximation in  a
relatively wide range in~$x$;
\item higher order corrections such as ${\cal O}(N^3\Lambda^3/m^3)$
which also appear in the exponent modify the shape of the spectrum near $x\sim 1$.
\end{itemize}

Regarding the non-perturbative fragmentation function as a set of
non-perturbative power corrections is useful only if the
perturbative sum itself is performed taking Sudakov logs as well
as infrared renormalons into account. This is the purpose of DGE.
The significance of Sudakov logs at large $x$ has long been
understood, while that of renormalons is more subtle. In
applications of Sudakov resummation in the asymptotic regime (e.g.
in the coefficient function for a single-particle inclusive cross
section at large $q^2$ -- see Sec.~4 and~5) power corrections are negligible, 
and
a fixed-logarithmic accuracy, e.g. NLL, is sufficient. For the
heavy-quark fragmentation function, on the other hand, power
corrections are {\em essential}, and therefore so is the proper
separation between the perturbative sum and the power corrections.
A proper separation can be achieved only if renormalons, which
manifest themselves as subleading Sudakov logs, are resummed.
Furthermore, we found that dealing with the renormalon ambiguity
by means of a PV prescription, one avoids Landau singularities and
obtains a sensible asymptotic behavior at large $N$.

The numerical analysis demonstrates the significance of
this property. In contrast with the fixed-order result or even the
NLL resummation, the DGE perturbative result with PV
regularization (see Fig.~\ref{fig:nlldgefull}), 
does not become large nor negative, but
remains instead positive up to $x = 1$. This is a result of {\sl
refining} the perturbative calculation for heavy-quark
fragmentation, by including the effects of leading as well as
subleading soft logarithms via DGE, while introducing a systematic
prescription (PV) to separate the perturbative and
non-perturbative contributions. Upon convolution with the properly
fitted non-perturbative contribution, this feature allows for a good
description of the data up to $x \simeq 1$.

While concentrating on the large-$x$ region, many other features of the
fragmentation process have been simplified. We neglected power
corrections in $m^2/q^2$ as well as perturbative corrections of
order $\alpha_s^2$ which are not logarithmically enhanced. We also
neglected power corrections of the type $\Lambda/m$ (not enhanced by
$N$), which
certainly play an important role at small $N$. All these aspects
definitely deserve further investigation in the future. In spite
of these simplifications and of the fact the non-perturbative
correction we employ has minimal flexibility, 
we obtained a fairly good description (Fig.~\ref{fig:shiftcurves} 
and~\ref{fig:curves}) of the ALEPH data on $B$-meson production. 
In doing so we extracted a value for $\alpha_s$ which is
compatible with other determinations (although a systematic error
analysis has not been performed here). We also found that the
non-perturbative parameters for the fragmentation function match
the expectations: the shift of the perturbative spectrum in $x$ is
indeed of order $\Lambda/m$ and $\epsilon_2$ in Eq.~(\ref{eq:np})
turns out to be very small.

\vskip 30pt
\noindent{\large {\bf Acknowledgments}} \\

We would like to thank Eli Ben-Haim, Volodya Braun, Stefano
Catani, Alberto Guffanti, Gregory Korchemsky, Chris Sachrajda, Gavin Salam,
George Sterman and Bryan Webber for very useful discussions. EG
thanks the DFG for financial support.
MC thanks the CERN Theory Division for hospitality and support.\\

\end{document}